\documentclass[usenatbib,referee]{mn2e}
\usepackage{xcolor}\usepackage{graphicx}
\def\l{$\lambda$}
\def\mbh{$M_{\rm BH}$\/}
\def\nh{$n_{\mathrm{H}}$\/}
\def\lledd{$L/L_{\rm Edd}$}

\def\rfe{$R_{\rm FeII}$}

\def\feiiq{\rm Fe{\sc ii}$\lambda$4570\/}

\def\ltsima{$\; \buildrel < \over \sim \;$}
\def\ltsim{\lower.5ex\hbox{\ltsima}}  
\def\gtsima{$\; \buildrel > \over \sim \;$}

\def\gtsim{\lower.5ex\hbox{\gtsima}} 
\def\h{$H_{0}$}

\def\civ{{\sc{Civ}}$\lambda$1549\/}

\def\cmq{cm$^{-2}$\/}
\def\cm3{cm$^{-3}$\/}
\def\hb{{\sc{H}}$\beta$\/}

\def\hbnc{{\sc{H}}$\beta_{\rm NC}$\/}
\def\mgii{{Mg\sc{ii}}$\lambda$2800\/}

\def\ciii{{\sc{Ciii]}}$\lambda$1909\/}
\def\oiiiopt{{\sc{[Oiii]}}$\lambda\lambda$4959,5007\/}
\def\o4363{{\sc{[Oiii]}}$\lambda$4363\/}

\def\siiii{Si{\sc iii]}$\lambda$1892\/}
\def\aliii{Al{\sc iii}$\lambda$1860\/}

\def\feiiopt{{Fe \sc{ii}}$_{\rm opt}$\/}
\def\feii{{Fe\sc{ii}}\/}
\def\siii{{Si\sc{ii}}$\lambda$1814\/}
\def\feiii{{Fe\sc{iii}}\/}
\def\fe{{\sc{Fe}}\/}

\def\vr{{$v_{\mathrm r}$}}

\def\fe76087{{\sc [Fe vii]}$\lambda$6087\/}
\def\oiii{{\sc [Oiii]}$\lambda$5007}

\def\kms{km~s$^{-1}$}
\def\kmpc{km~s$^{-1}$\/ Mpc$^{-1}$\/}

\def\ergss{ergs s$^{-1}$\/}

\def\om{$\Omega_\mathrm{M}$}
\def\ol{$\Omega_{\Lambda}$}
\def\ok{$\Omega_\mathrm{k}$}
\def\apj{ApJ}
\def\aj{AJ}
\def\apjl{ApJL}
\def\apjs{ApJS}
\def\araa{ARAAp}
\def\nat{Nature}
\def\mnras{MNRAS}
\def\aap{AAp}

\def\pasp{PASP}
\def\pasj{PASJ}
\def\memsai{MemSAIt}
\usepackage{enumerate}

\title{Highly Accreting Quasars: Sample Definition and { Possible } Cosmological Implications}
\begin{document}
%% one line. H}

\author[P. Marziani \& J. W. Sulentic]{P. Marziani$^{1,2}$\thanks{E-mail:
paola.marziani@oapd.inaf.it}, J. W. Sulentic$^{2}$\\
$^{1}$INAF, Osservatorio Astronomico di Padova, Vicolo dell' Osservatorio 5, IT 35122, Padova, Italy\\
$^{2}$Instituto de Astrof\'\i sica de Andaluc\'\i a (CSIC), C/ Camino Bajo de Hu\'etor 50, 18008 Granada, Spain}
%\altaffiltext{1}{IAA-CSIC, Granada, Spain}
%\altaffiltext{2}{INAF, Osservatorio Astronomico di Padova, Italy. e-mail: paola.marziani@oapd.inaf.it}
\maketitle
%\altaffiltext{1}{Sabbatical Visitor, IAA-CSIC}


\begin{abstract}
{ We propose a method to identify quasars radiating closest to the Eddington limit, defining
 primary and secondary selection criteria in the optical, UV and X-ray spectral range based on the 4D eigenvector 1 formalism. We then show that it is possible to derive a redshift-independent estimate of luminosity for extreme Eddington ratio sources.  Using preliminary samples of these sources in three redshift intervals  (as well as 
two mock samples), we test a range of cosmological models.  Results  are consistent with 
concordance cosmology but  the data are insufficient for deriving strong  constraints. 
Mock samples indicate that application of the method proposed in this paper using dedicated observations 
would allow to set stringent limits  on \om\ and significant constraints on \ol.}
%More than simple empirical descriptors, the 4D eigenvector 1 parameters are all consistent with a 
%restricted range of physical conditions in these extreme Eddington ratio sources. 
%They can be effectively selected using the  principal optical, UV and X-ray parameters in the 4D eigenvector 1 formalism. 
\end{abstract}




\section{Introduction}
\label{intro}

The concordance cosmology model \citep{spergeletal03,riessetal09,komatsuetal11}
favors a flat Universe (\om+ \ol=1) and significant energy density associated with a 
cosmological constant (\ol=0.72). Inference of a significantly non-zero \ol\ rests on the 
use of supernov\ae\ as standard candles \citep{perlmutteretal97,perlmutteretal99} and 
studies of the cosmic microwave background  radiation \citep[e.g.][]{tegmarketal04}.
The most recent Wilkinson Microwave Anisotropy Probe (WMAP) results indicate that the baryonic acoustic oscillation scale,  the
Hubble constant and  the densities are determined to a precision of 
$\approx$1.5\%\ \citep{hinshawetal12}.  Even if the large-scale distribution of galaxies
\citep{schlegeletal11,nuzaetal12} and galaxy clusters  \citep[e.g.,][]{allenetal11} provide 
additional constraints, it is urgent that independent lines of investigation  are devised to test 
these results. It remains important  because \om\ is not very tightly constrained by supernova surveys \citep{conleyetal11}: \om $\ltsim$0.5 at 2-$\sigma$\ confidence level from supernov\ae\ mainly at $z \ltsim$ 1.5 \citep{campbelletal13}.  Little information on \om\ has been extracted from the redshift range $1 \ltsim z \ltsim 3$, where \om\ is most strongly affecting the metric. In addition, recently announced cosmological parameter values from the Planck-only best-fit 6-parameter $\Lambda$-cold dark matter  model differ from the  previous WMAP estimate, and yield  \ol=0.69$\pm$0.01  \citep[][]{planck13}, significantly different from  \ol = 0.72, \om =0.28 of the concordance cosmology adopted in the past few years \citep{hinshawetal09}.  

It is perhaps appropriate that we reconsider the cosmological  
utility of quasars  at the fiftieth anniversary of their discovery \citep[e.g., ][]{donofrioetal12}. 
It is well known that use of quasar properties  for independent  measurement of cosmological parameters is fraught with difficulties.
Quasars show properties that make them potential cosmological probes (see e.g. 
\citealt{bartelmannetal09}): they are plentiful, very luminous, and detected at early 
cosmic epochs (currently out to $z \approx 7$, \citealt{mortlocketal11}). The downside 
is that they show a more than 6dex spread in luminosity and are also anisotropic
radiators. Quasars are thought to be the observational manifestation of accretion 
onto super-massive black holes. Accretion phenomena in the Universe show a scale 
invariance with respect to mass and, indeed, we observe similar quasar spectra 
over the entire luminosity/mass range. It is not surprising that pre-1990 quasar 
research tacitly assumed that high and low luminosity active galactic nuclei (AGN) were
spectroscopically similar as well. The key to possible cosmological utility lies in realizing that 
quasars show a wide dispersion in observational properties  which is a reflection  of  
the source Eddington ratio. 

\subsection{Quasar Systematics}

\begin{figure}
%\epsscale{.50}
\includegraphics[scale=0.5]{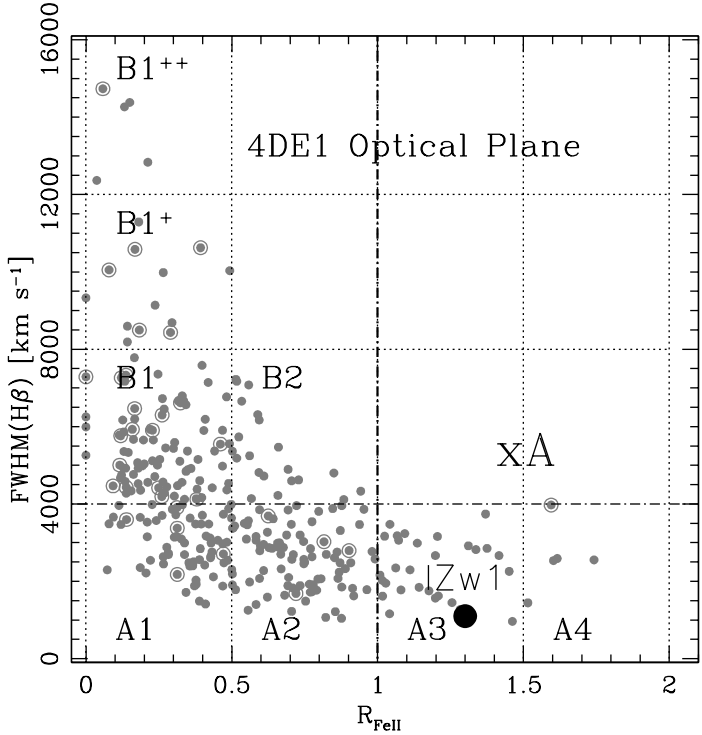}
\caption{The optical plane of the 4DE1 space, FWHM \hb\ vs \rfe. Data points (in grey) are from the sample of \citet{zamfiretal10}; circled points represent RL sources. The plane is binned following \citet{sulenticetal02} to identify spectral types (thin dot-dashed lines). The thick dot-dashed line separates extreme Pop. A source by the criterion \rfe$\ge$1.0.  The large filled circle identifies the extreme Pop. A source I Zw 1.  \label{fig:e1}}
\end{figure}

The goal of systematizing observational properties of quasars has made  considerable 
progress in the past 20 years and this makes possible  discrimination of sources by 
$L/L_\mathrm{Edd}$. An important first step involved a principal component analysis 
of line profile measures from high S/N spectra of 87  PG quasars \citep{borosongreen92}.
That study revealed  of Eigenvector 1 (E1) correlates including a trend of 
increasing optical \feii\ emission strength with decreasing FWHM H$\beta$\ and peak 
intensity of \oiii. Source luminosity was found to be part of the second Eigenvector and is 
therefore not directly correlated with the Eigenvector 1 parameters. A second step in 
quasar systematization attempted to identify more key line and continuum diagnostic 
measures.  4D Eigenvector 1    
\citep[4DE1,][]{sulenticetal00a} included  two E1 broad line measures: 1) full width half maximum of 
broad H$\beta$\ (FWHM H$\beta$) and 2) optical \feii\ strength { defined by the equivalent width W or by the intensity ratio }
\rfe= W(\feiiq)/W(H$\beta$) $\approx$ $I$(\feiiq)/$I$(H$\beta$), where \feiiq\ indicates the blend of \feii\ emission between 4434 \AA\ and 4684 \AA. Figure \ref{fig:e1} shows the optical plane of 4DE1 as defined by the  470 brightest quasars from SDSS DR5. Domain space in the figure is binned in such a way that all 
sources within a bin are the same within measurement uncertainties, and can be assigned a well-defined spectral type \citep{sulenticetal02,zamfiretal10}.
The majority of quasars occupy bins A2 and B1 with tails extending towards bins with sources 
showing stronger \feii\ emission or broader H$\beta$ profiles. Currently  eight bins in intervals 
of FWHM = 4000 \kms\ and \rfe = 0.5 are needed  to fully map source occupation. 4DE1 added two additional principal parameters: 3) a measure discussed in \citet{wangetal96}
of the soft X-ray photon index ($\Gamma_\mathrm{soft}$) and 4) a measure of the \civ\ broad 
line profile shift  \citep[at half maximum, ][]{sulenticetal07}. Points of departure from BG92 
involve: a) removal of \oiii\ measures as 4DE1 correlates, b) consideration 
of differences in parameter space occupation between radio-quiet (RQ) and radio-loud (RL) 
sources \citep{sulenticetal03,zamfiretal08} as well as c) division of  sources into two 
populations (A and B) designed to  emphasize source spectroscopic differences.  

Population A 
sources show FWHM H$\beta <$4000 \kms,  stronger \rfe, a soft X-ray excess and \civ\ 
blueshift/asymmetry. Pop. A includes sources often called Narrow line Seyfert 1s (NLSy1s).  
The H$\beta$ broad-line profile shows an unshifted Lorentzian shape
\citep{veroncettyetal01}.   Population B sources show FWHM H$\beta >$4000 \kms\ with weaker \rfe, no soft  X-ray excess and usually  no \civ\ blueshift.  Most RL sources are found in the 
Pop. B domain of Figure \ref{fig:e1}. Pop. B sources usually require a double Gaussian model  
(broad unshifted + very broad redshifted components) to describe \hb. Labels on the bins shown 
in Figure \ref{fig:e1} identify regions of occupation for the two source populations following the spectral type classification of \citet{sulenticetal02}. Other important 
4DE1 correlates  exist. 
%and we will consider some in connection with the selection 
%of cosmologically useful quasars. 
The present 4DE1 parameters represent those for which large 
numbers of reasonably accurate measures are available and for which clear 
correlations can be seen. 

4DE1 parameters  measure: FWHM H$\beta$ -- the dispersion in low ionization line 
broad line region (BLR) gas  velocity (it is the virial estimator of choice at low $z$),  \rfe\ -- the relative strengths of 
\feii\ and \hb\ emission--likely  driven by density \nh, ionization and metallicity, $\Gamma_\mathrm{soft}$ -- the strength of a soft 
X-ray excess viewed as a thermal signature of accretion \citep{shields78,malkansargent82}
and \civ\ shift  -- the amplitude of systemic 
radial motions in high ionization BLR gas, possibly due to an accretion   disk wind \citep{koniglkartje94,proga03,konigl06}. 
%Going 
%beyond what physical insights each measure might give us 
If we ask what might drive the source 
distribution in Figure \ref{fig:e1} (or any of the other planes of 4DE1) ,
%especially since source luminosity is not the driver. 
the answer is most likely Eddington ratio. The idea that Eddington ratio drives 
4DE1 goes back to the first E1 study  \citep{borosongreen92} and  has received  considerable support  in the past 20 years \citep[e.g.,][]{boroson02,marzianietal03b,baskinlaor04,grupe04, yipetal04,aietal10,wangetal11,matsuoka12,xuetal12}. Black hole mass, and orientation are sources of scatter in the 4DE1 sequence. The value  \lledd\ $\approx$ 0.2$\pm$0.1 corresponds to the boundary between Pop.  A and B sources  \citep[for a black hole mass $\log$ \mbh\ $\sim$ 8.0, ][]{marzianietal01,marzianietal03b}, and   may be ultimately related to a transition between a geometrically thin, optically thick disk and an advection dominated, ``slim'' disk \citep[e.g., ][]{rozanskaczerny00,collinetal02,chenwang04}.  Identification of the most 
extreme accretors (potentially ``Eddington standard candles'') offers the best hope toward finding a 
cosmologically useful sample of quasars. 

\subsection{Finding  ``Eddington Standard Candles''}

Any  attempt to use quasars as redshift-independent distance estimators must be tied to identification 
of a special class with some well-defined observational properties. These properties should be chosen 
to provide an easy link to a physical parameter related to source luminosity. If \lledd\ is known then
standard assumptions can lead to a $z$-independent estimate of source luminosity  
since source luminosity estimation is connected to estimation of \mbh\ (\S \ref{lumv}). 
%While most present-day  estimates of  \lledd\ are rather rough,  

Empirical studies  show that the \lledd\ distribution truncates near \lledd\ $\approx$1 \citep[e.g., ][]{woourry02,shenetal11}. We observe a few super--Eddington outliers but their masses have likely been significantly  underestimated
due to a special line-of-sight orientation (e.g. sources oriented face-on where virial motions make little or no  contribution to line width \citealt{marzianietal03b} and \citealt{marzianisulentic12}). The range of \lledd\ is $0.01 \la $\lledd $\la 1$ for luminous Seyfert 1 nuclei and 
quasars \citep{woourry02,marzianietal03b,steinhardtelvis10,shenkelly12} with a
much broader range if low-luminosity sources (of less  interest for the present study) are included. 
The flux-limited distribution of \lledd\ is subject to a strong, mass dependent Malmquist 
bias \citep{shenkelly12} that may explain claims of a significantly narrower range. The 
minimum detectable \lledd\ for a limiting magnitude $m_\mathrm{B,lim}$\ can be   written as
$L/L_{\rm{Edd,min}} \propto \frac{1}{M_{\rm{BH}}}
10^{-0.4 m_\mathrm{B,lim}} f(z)^2 (1+z)^{(1-a)}$, where $f(z)$\ is the
redshift function appearing in the comoving distance definition, and $a$\
is the visual continuum spectral index. The above  expression shows that only lower
\lledd\ values are lost at increasing redshift and that the effect is \mbh-dependent.   

{ A physical motivation underlies our search for sources radiating at
extreme Eddington ratio.} When accretion becomes super-Eddington, emitted
radiation is advected toward the black hole, so that the source luminosity
{ tends to saturate  if accretion rate $\dot{m} \gg
\dot{m}_\mathrm{Edd}$, where $\dot{m}_\mathrm{Edd}= L_\mathrm{Edd}/ \eta
c^2$\ is the Eddington accretion rate for fixed efficiency. { The parameter $\eta$\ can be assumed  $\approx
\frac{1}{6}$, as for $\alpha$ accretion disks \citep[e.g.,][]{shakurasunyaev73}, or    equal 1 \citep{mineshigeetal00}. } }  Radiative efficiency is expected to decrease with increasing Eddington ratio and luminosity to
increase with log accretion rate
\citep{abramowiczetal88,szuszkiewiczetal96,collinkawaguchi04,kawaguchietal04}.
{ Current models indicate a saturation value of a few times the
Eddington luminosity for the bolometric luminosity
\citep{mineshigeetal00,wataraietal00}. The observed \lledd\
distribution of high Eddington ratio candidates considered in this paper
is consistent with a limiting \lledd $\rightarrow 1$ (\S \ref{identif}), as found in earlier 
 empirical  \citep[e.g.,][]{marzianietal03b}.}  

The primary goal of this paper is to find sources radiating near the limit  \lledd\ \ 
using emission line intensity ratios. These sources should be the
quasars most easily found in flux-limited surveys i.e., the bias described
above will only result  in selective loss of quasars radiating at lower Eddington ratios for fixed \mbh.  
We suggest that the 4DE1 formalism provides a set of parameters currently best 
suited to identify extreme Eddington radiators (\S \ref{sample}). 
%Not all are useful at the present time for practical 
%reasons but they hold promise for future refinements of a search for standard Eddington candles.
A second goal   is to propose a  method  to for using high \lledd\ quasars as redshift-independent luminosity estimators  where the constant parameter is Eddington ratio rather  than luminosity (\S \ref{lumv}). We  report { explorative calculations} and preliminary results (\S \ref{cosmo}) and discuss possible improvements (\S \ref{mock}) along with limits and uncertainties (\S \ref{seleffects}).

\section{Sample Selection and Measurements}
\label{sample}

\subsection{I Zw 1 as a prototype of highly accreting quasars}
 
 The much studied  NLSy1  source I Zw 1 is considered to be a low-$z$\  ($z \approx 0.0605$) prototype of  quasars radiating close  to the Eddington limit. It is located in bin A3 of Figure \ref{fig:e1}.  The 4DE1 properties of I Zw 1 are:

\begin{enumerate}
\item FWHM H$\beta$= 1200$\pm$50 \kms\ \citep{borosongreen92};
\item \rfe = 1.3$\pm$0.1 \citep{borosongreen92};
\item \civ\ blueshift at FWHM (relative to rest frame) $\Delta$\ \vr =  -1670$\pm$100 \kms \citep{sulenticetal07};
\item soft-X photon index $\Gamma_\mathrm{soft}$= 3.050 $\pm$ 0.014 \citep{wangetal96}.
\end{enumerate}

All 4DE1 parameters for I Zw 1 are extreme making it a candidate for the sources we are seeking. 
Conventional \mbh\ and \lledd\ estimates for this source  yield  $\log$ \mbh $\approx$ 7.3-7.5 in solar units 
and  $\log $\lledd $ \approx$ -0.13 to +0.10 \citep{vestergaardpeterson06,assefetal11,negreteetal12,trakhtenbrotnetzer12}.
%$\log $\lledd\ is  consistent with 0 and  $\log$ \mbh $\approx$ 7.3 in solar units. 
 %\lledd\ estimates used two different bolometric corrections.  \mbh\ estimates are not fully independent. 
 If  we consider the  average \mbh\ of \citet{vestergaardpeterson06} and \citet{trakhtenbrotnetzer12}  adopting   two different bolometric corrections  ($L$ = 10 $\lambda L_{\lambda}$(5100) and the $B.C.$ of  \citealt{nemmenbrotherton10}) we obtain $\log$\ \lledd $\approx -0.11 \pm 0.17$, consistent with unity.  

We need a statistically useful sample  of extreme sources for any attempt at cosmological application. It is premature to discuss here all the corrections that must be applied; however, we seek sources similar to, or more extreme than I Zw 1. 
%At this point in time we must rely on \rfe\ to select extreme sources at low redshift.  

FWHM H$\beta$ will be the least useful 4DE1 parameter because all Population A 
sources -- 50\% of the low $z$\ quasar population -- show  FWHM H$\beta$ $<$4000 \kms, and 
only a small fraction of this population is likely to involve extreme  Eddington radiators 
based upon current \lledd\ estimates  \citep{marzianietal03b}. This is also true for higher $z$\ sources.
In addition,  FWHM of \hb\ increases slowly but systematically with $L$, and there is a minimum  FWHM possible if gas is moving virially and \lledd\ $\le$ 1 \citep{marzianietal09} that is $\approx $ 4000 \kms\ at $\log L \sim 48$ [\ergss]. Currently  only very high $L$ sources can  be studied with any accuracy at high $z$. Therefore, we will relax any FWHM limit introduced in the definition of spectral types in  low redshift samples ($z \ltsim 0.7$). This concerns only spectral types A3 and A4 for which there is no danger of confusion with broader sources, as shown by Fig. \ref{fig:e1}. The spectral types based only on \rfe\ will be indicated with A3$^\mathrm{m}$\ (1 $\le $ \rfe$<$ 1.5) and A4$^\mathrm{m}$ (1.5 $\le $ \rfe$<$ 2.0), or together, with  ``xA'' (\rfe$\ge$ 1.0). 

At this stage we adopt \rfe$>$1.0 as a primary selector of low redshift extreme Eddington 
radiators which are by definition Pop. A sources. Clearly significant numbers of such candidates can be identified, $\approx$ 10\%\ of all low-$z$\ quasars (\citealt{zamfiretal10}, \citealt[c.f.][]{shenetal11}). In a  recent analysis of \mgii\ line profiles for an SDSS-based sample of 680 quasars ($z$\ in the range 0.4 -- 0.75) we found $n$ = 58 candidate A3$^\mathrm{m}$/A4$^\mathrm{m}$\ sources. They are high confidence candidates because  spectral S/N is high enough to be certain about the extreme \feii\ emission. $\Gamma_\mathrm{soft}$ 	\  and \civ\ measures exist for 
almost none of these sources at this time  but most PG candidates at lower redshift ($z <$0.5) show extreme 
\civ\ and soft X-ray properties \citep{borosongreen92,wangetal96,sulenticetal07}.

The left panel of Figure \ref{fig:examphb} shows profile model fitting to the median bin A3 spectrum \citep[sample of][]{marzianietal13a}. The \rfe\ value is consistent with the one of I Zw 1 reported  in \citet{negreteetal12}. 
%The H$\beta$\ fit yields FWHM which can provide an \mbh\ estimate (and \lledd) using  standard procedures.  
Details of model fitting procedures can be found in \S\  \ref{meas} and  \citet{marzianietal10}.

%There is little doubt about  the bin A4 assignment of such sources  given the very strong FeII, and weak H$\beta$, emission. 
%These measures, coupled with the assumption that \lledd $\approx$1, yield a candidate standard candle.

\subsection{Selecting High $z$\ Samples of Eddington Candles}
\label{highz}

Without IR spectra we lose the \rfe\ selector employed at $z <$1.0.  We must take advantage of quasar abundant rest-frame UV  spectra.  The first choice 
would be to use measures of \civ\ shift, asymmetry and equivalent width because extreme sources tend to show  low EW \civ\ profiles that are blueshifted and blue asymmetric. A measure of \civ\ shift was chosen as a 4DE1 parameter because it is the best measure of extreme \civ\ properties. At low $z$\ we rely on satellite observations where the HST archive provides the highest S/N and resolution UV spectra for our low $z$\ candidates. Data exist in the HST archive for about 130 low $z$\ sources of which $n$ = 11 are bin A3  -- A4\ sources (\rfe$>$1.0).  Ten of the eleven \feii\ extreme sources show a significant \civ\ blueshift confirming its utility as an extreme Eddington selector \citep{sulenticetal07}. The majority of Pop. A sources with \rfe$<$1.0 and Pop. B sources show a smaller blueshift and in many cases zero shift or a redshift ($n$ = 3 with shift $>$ 10$^3$ \kms).  

However, we do not know  the geometry of the outflows implied by the profile blueshift. We may miss a considerable number of candidates where orientation diminishes the blueshift. There is also the issue of whether \civ\ blueshifts are quasi-ubiquitous in high $z$\ quasars \citep{richardsetal11}. If \civ\ blueshifts are more common at high $z$\ then they may not be useful as a clearcut selector. A further problem is that a selector involving a line shift requires a more 
accurate estimate of the quasar rest frame which is often missing in high redshift sources.

%Both Pop. A and B  sources have also been identified at high $z$\ \citep{marzianietal09} although  sources show broader lines  because they are much more massive than our low $z$\  quasars (if no super-Eddington radiators exist; see Figure 11 in \citealt{marzianietal09}). Aside from showing I Zw 1 as an example of a low $z$\ extreme source we can use it to help us establish 
%the most effective selector for high $z$\ samples.  The first choice is obviously \civ. Our low redshift sample \citep{bachevetal04,sulenticetal07} shows \civ\ blueshifts only in Pop. A sources ($\sim$50\%) and extreme values in only 10-20\%\ of Pop. A sources.  
 
\begin{figure}
%\epsscale{.4750}
\includegraphics[scale=0.35]{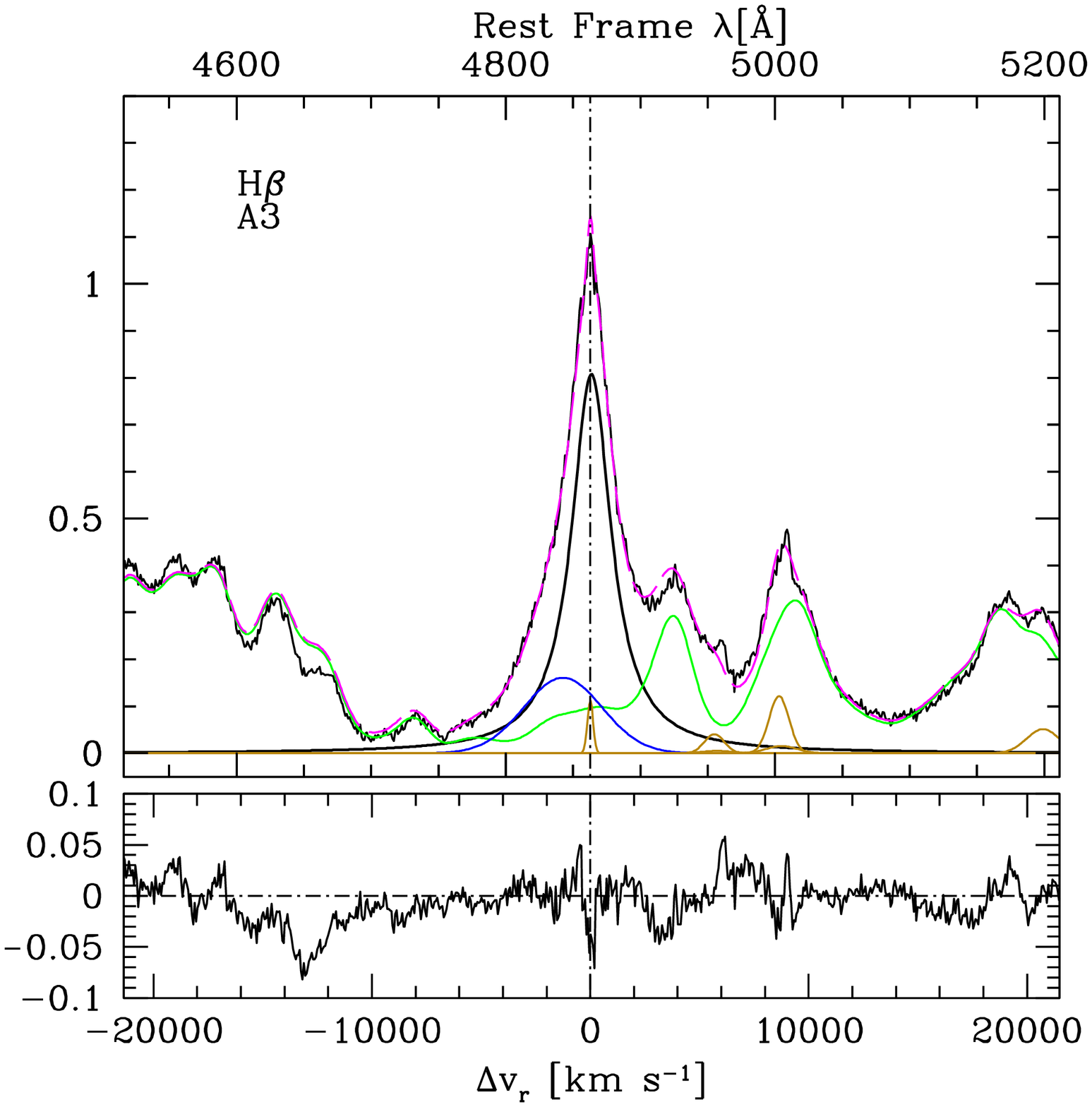}
\includegraphics[scale=0.35]{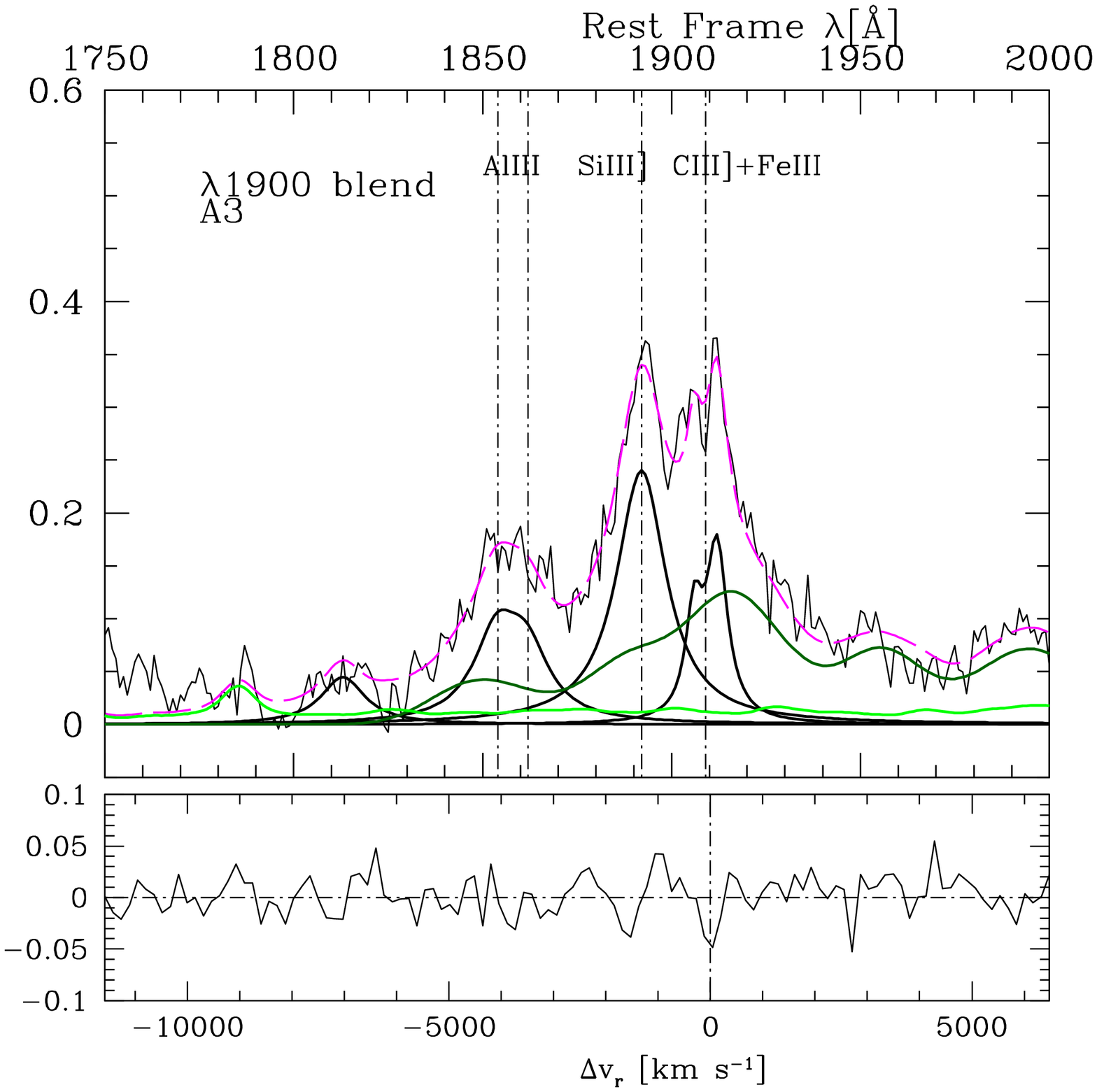}
\caption{Broad emission line fits of the \hb\ spectral region (left) and \aliii\ for  median spectra of two A3 samples, shown after continuum subtraction. Abscissa is rest-frame wavelength; ordinate is normalized intensity. The dashed magenta line shows the model spectrum. The lower panel shows the difference between the observed and the model spectra.   The lemon green lines  trace  \feii\  and the dark green one \feiii\ emission in correspondence of the 1900 blend. The \hb\ broad profile (thick line) is isolated after \feii\  and narrow line subtraction. Only a faint \hb\ narrow component is detected (golden line). The blue line shows the blueshifted residual associated to a non-virial \hb\ component.  In the right panel, the thick black lines on the right panel trace the \siii, \aliii, \siiii, and \ciii\ in order of increasing wavelength. Note that the \aliii\ profile is modelled as a  doublet with intensity ratio 1858/1863 = 1.2:1.   \label{fig:examphb}}

\end{figure}

We propose an alternative high $z$\ selector involving the 1900 emission line blend of  \aliii,  \siiii\ and \ciii. The blend involving these lines constrains the physical conditions in the broad 
line emitting gas much the same way as measures of very strong optical \feii. The definition of this 
selector comes from three sources: 1) measures from a composite spectrum of HST archival data for 
10 bin A3 sources \citep[right panel of Fig. \ref{fig:examphb}, ][]{bachevetal04}, 2) line ratio  measures for I Zw 1 (our bin A3 prototype) from a high S/N and resolution HST spectrum  \citep{laoretal97a,negreteetal12} and 3) measures of SDSS1201+0116 
(a high $z$\ bin A4 analogue of I~Zw~1 from a high S/N VLT spectrum \citealt{negreteetal12}).
  
\citet{bachevetal04} noted that the \siiii/\ciii\ intensity ratio decreased monotonically 
along the 4DE1 sequence by a factor  $\approx$4 from bin A3 ($\approx$0.8)  to bin B1+ ($\approx$0.2). This implies a selection constraint for extreme sources \siiii/\ciii\ $>$0.6-0.7. Measures using very high S/N spectra and on  the 1900 blend of the A3 composite spectrum (Fig. \ref{fig:examphb}) yield a  selection 
criterion based on two related ratios:
\begin{enumerate}
\item \aliii/\siiii/ $\ge$ 0.5  and 
\item \siiii/\ciii\ $\ge$1.0.
\end{enumerate}
Both our low and high redshift selectors are based on more than empiricism because
they effectively constrain the range of physical conditions in the line emitting region 
of the extreme quasars \citep{baldwinetal96,baldwinetal04,marzianietal10,negreteetal12}. The extreme \feii\ emission from these sources has 
been discussed in terms of the densest broad line  emitting region (\nh $\sim$10$^{12}$  \cm3;
\citealt{negreteetal12}) and extreme metallicity again plausibly connected with high accretion rates.

%\footnote{Not all Pop. A --which include all NLSy1 sources--{ show } the  same 
%properties as I Zw 1.} 

\subsection{Identification of Preliminary Samples}
\label{identif}
 
Equipped with  low and high redshift selection criteria we identify
three samples of extreme Eddington sources over the range $z =$ 0.4 -- 3.0. 

\begin{enumerate}
\item Sample 1: 58 sources with   \hb\ spectral coverage from an SDSS DR8 sample of 680 quasars 
in the range $z \approx 0.4 - 0.75$\  \citep{marzianietal13}. Their spectral bins are  A3 and A4. 
Table \ref{tab:hb} lists the 30 and 13 sources in bin A3 and A4  with  S/N $\ge$ 15 at 5100 \AA. Formats for Tables \ref{tab:hb} and \ref{tab:aliii}  are: ID, redshift, restframe flux $\lambda$f$_\lambda$, S/N of spectrum, FWHM of virial estimator used, FWHM uncertainty and sample ID. FWHM uncertainties for  sample 1 \hb\ profiles were assumed to be 10\%.

 \item Sample 2: 7 sources from a sample of 52 Hamburg-ESO quasars \citep{marzianietal09} in the range   $z$ = 1.0 -- 2.5 (all but two $1 \ltsim  z  \ltsim 1.6$) with high S/N (VLT-ISAAC) spectra  of the H$\beta$ region.  They all satisfy the criterion \rfe $\ge$ 1.0 within observational uncertainty; 3 of them are however borderline sources with \rfe $\approx$ 1.0.  The ISAAC sources are meant to cover a redshift range where \hb\ observations are very sparse.  

\item  Sample 3: 63 SDSS sources (additional  candidates were identified but require higher S/N data) are   listed in Table \ref{tab:aliii}.  We extracted spectra for $\approx$3000\ sources from SDSS DR6 with coverage  of the 1900 blend  ($2.0 \ltsim z\ \ltsim $ 2.6). Selected sources show emission line   ratios \aliii / \siiii\ and \ciii /\siiii\  satisfying our criterion. A further restriction was imposed by excluding  sources with low S/N  ($<$ 15) spectra  that might have biased FWHM measures  as clearly shown  by \citet{shenetal08}.  This yielded  a subsample 3 of 42 sources  with S/N$\ge$ 15 on the continuum at 1800 \AA. FWHM uncertainties were estimated from simulated spectra as a function of \aliii\ EW, FWHM and S/N. 
 % 15 and (3b) of 26 with S/N$\gtsim$ 20 
\end{enumerate}

Sample 1 and 2   are based on flux limited samples analysed in the studies cited above. Sample 3 has no pretence of completeness: the lower fraction of identified xA sources is due to the low S/N of most spectra.  

\subsection{Emission Line Measurements }
\label{meas}

\begin{figure}
%\epsscale{.50}
\includegraphics[scale=0.35]{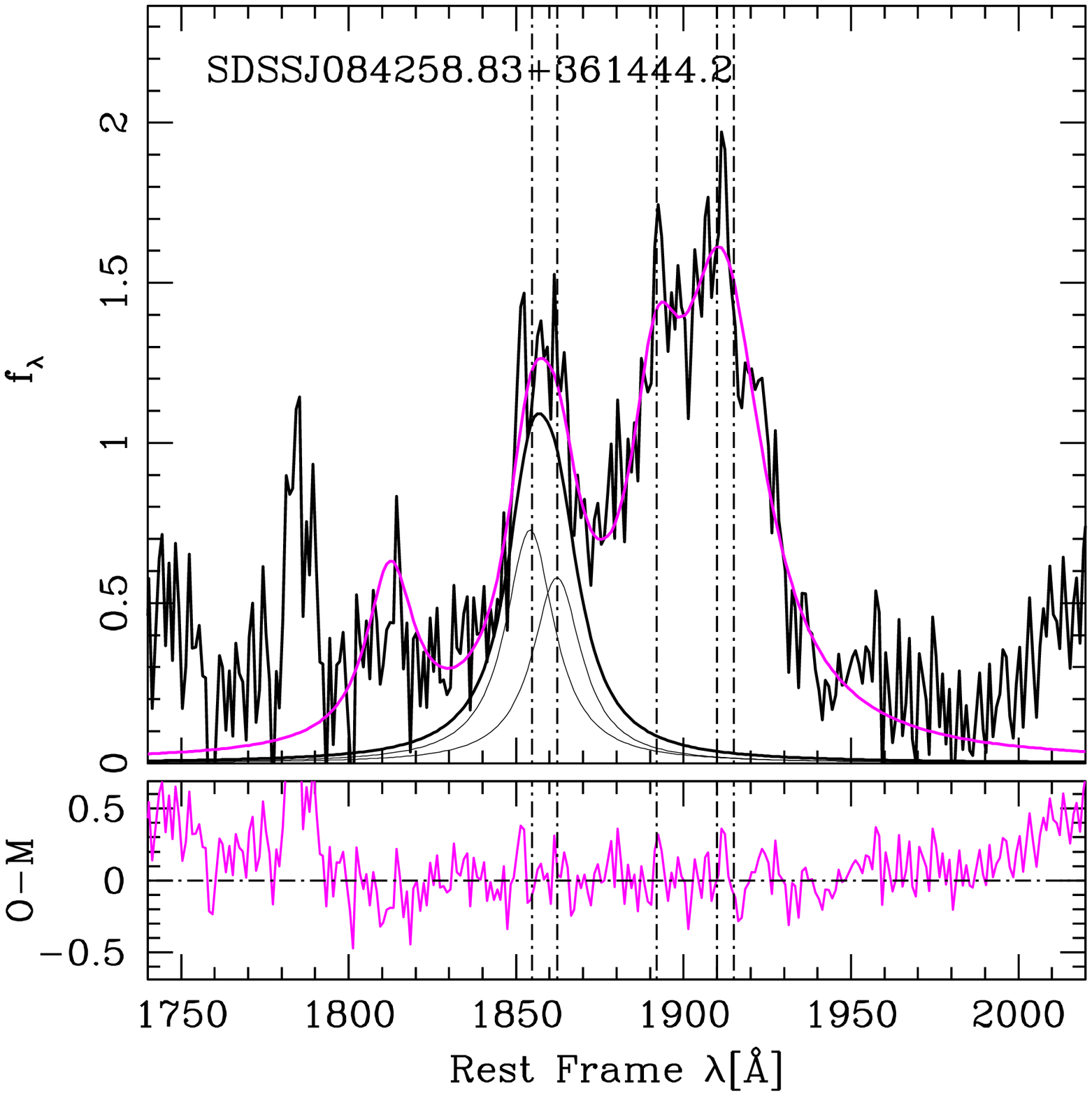}
\includegraphics[scale=0.35]{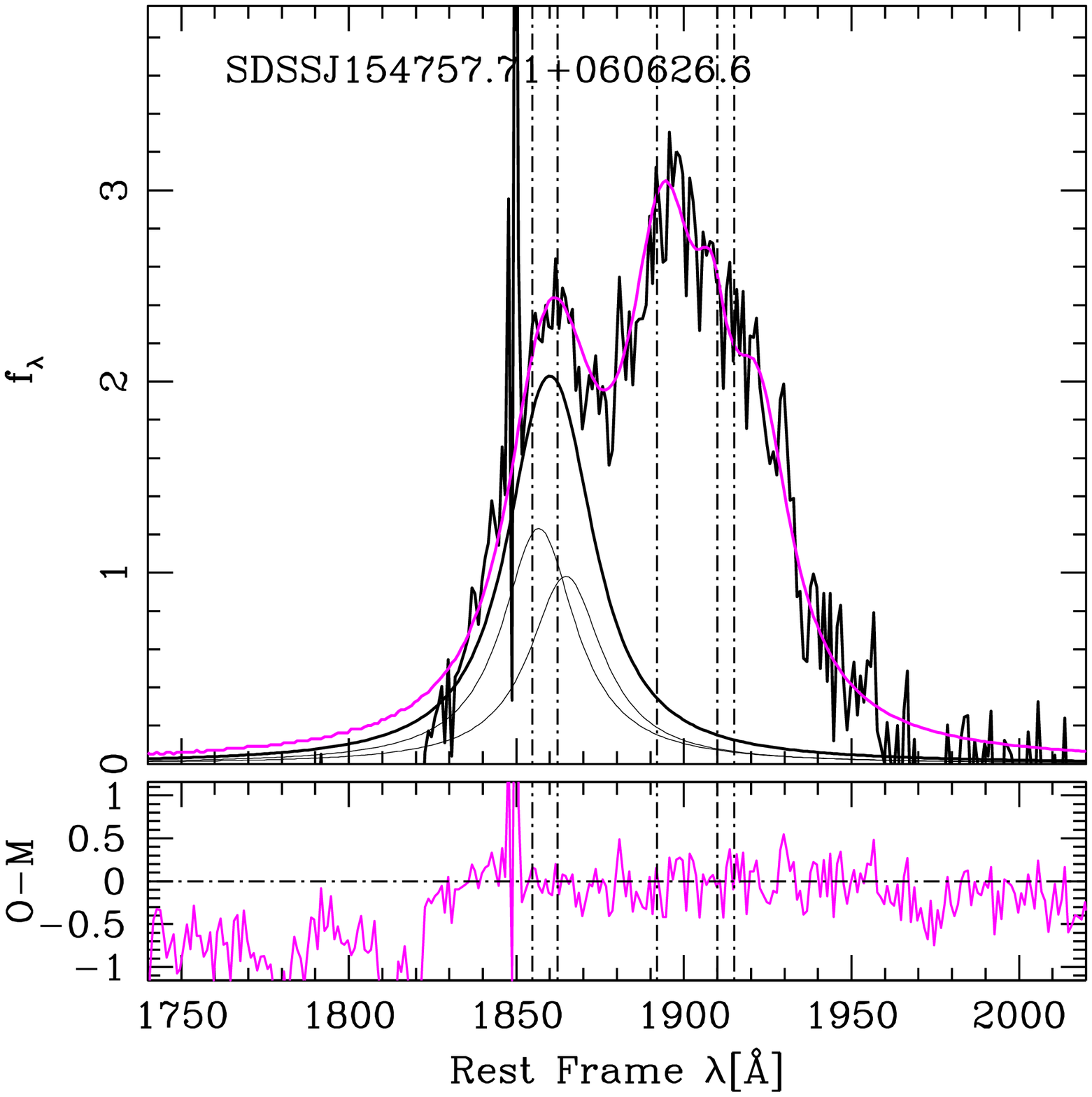}
\caption{Example of broad emission line fits of the \l 1900 \AA\ blend, shown after continuum subtraction. Abscissa is rest-frame wavelength; ordinate is specific flux  in units of 10$^{-15}$\ergss\ \cmq\ \AA$^{-1}$. The FWHM of the individual components of the \aliii\ doublet (thin lines) provide a virial broadening estimator. Total \aliii\ emission (thick line) can be modeled since the doublet is well-separated from \siiii\ and the rest of the blend. The magenta dashed line shows the full model of the  \l 1900 \AA\ blend which includes \siiii, and \ciii\ and \feiii\ emission. \label{fig:exampaliii} }

\end{figure}

Line ratios  yield 4DE1 bin assignments. In the optical,   \rfe\ is retrieved from the intensity of the Lorentzian profile representing \hb\ and from \feiiopt\ flux in the integrated over the wavelength range 4434 -- 4684 \AA\ \citep{borosongreen92,marzianietal03a}. In the UV, it is  especially the \aliii/\siiii\ ratio\ that, with or without \civ\ measures, provides a selector for high $z$\ extreme sources. Figure \ref{fig:exampaliii} shows model fits to the 1900 blends for two high $z$\ extreme source candidates. 

{ Measurements were made using a nonlinear multicomponent fitting
routine that seeks $\chi^2$\ minimisation between observed and model
spectra, i.e., {\sc specfit} incorporated into {\sc iraf} \citep{kriss94}.
The  procedure allows for simultaneous fit of continuum, blended iron
emission, and all narrow and broad lines identified in the spectral range
under scrutiny. Singly and doubly ionised iron emission (the latter
present only in the UV) were modelled with the template method. For the
optical \feii\ emission, we considered the semi-empirical template used by
\citet{marzianietal09}. This template was obtained from a high resolution spectrum of I
Zw 1, with  a model of \feii\ emission   computed by a photoionization code  in the 
wavelength range underlying \hb. For the UV \feii\ and \feiii\ emission, we
considered templates provided by \citet{bruhweilerverner08} and
\citet{vestergaardwilkes01}, respectively. The fitting routine scaled and
broadened the original templates to reproduce the observed emission. Broad
\hb\ and the 1900 \AA\ blend lines were  fit assuming a Lorentzian function. This
assumption follows from analysis of large samples of Pop. A sources
\citep{veroncettyetal01,marzianietal03b,zamfiretal10,marzianietal13a}:  
\citet{negreteetal13} verified that  H$\beta$, \aliii\ and \siiii\ can be fit with a Lorentzian function whose width is the same for the three lines in Pop. A sources studied with reverberation mapping. %\aliii, \siiii\ and \ciii\ peak wavelengths were also found to be suitable as   rest frame indicators.
%Lorentz function also provides a good fit to lines in the 1900 \AA\ blend
%as shown in the present work and by \citet{negreteetal13}. 
The broad \hb\ line profile is often asymmetric toward the blue. The line profile has been modeled adding to the Lorentzian component  an additional, shifted line component associated to outflows \citep[c.f., ][]{marzianietal13a}. 
Narrow lines (\oiiiopt\ and \hbnc) were fit with Gaussian functions.
% and also serve as  estimators of the local source rest frame. 
%Either FWHM \siiii\ or FWHM \aliii\ provide virial estimators consistent with FWHM H$\beta$\ used at lower $z$.  %In the UV,     which increases the accuracy of the \civ\ shift measures. 

The uncertainty associated with  FWHM\hb\ was 10\%\ for sample 1,
and as reported in \citet{marzianietal09} for the sample 2 (usually $\la $10\%).  For the UV virial estimator i.e., in cases of  lower S/N , synthetic blend profiles were computed as a function of S/N,
equivalent width, and individual line width for typical values in our
sample (e.g., S/N = 14,22; W(\aliii) = 5,10 \AA; FWHM = 3000,4000 \kms).
Statistical measurement errors on each source FWHM were assigned to the
value computed for the synthetic case with closest lower S/N, closest lower 
equivalent width and closest FWHM.
 %H$\beta$\ retains much value as the virial estimator of choice for low $z$  quasars -- and high $z$\ as well -- if IR spectra are available. 
 
 }

\subsection{Consistency of criteria}
\label{consist}

\begin{figure}
%\epsscale{.50}
\includegraphics[scale=0.5]{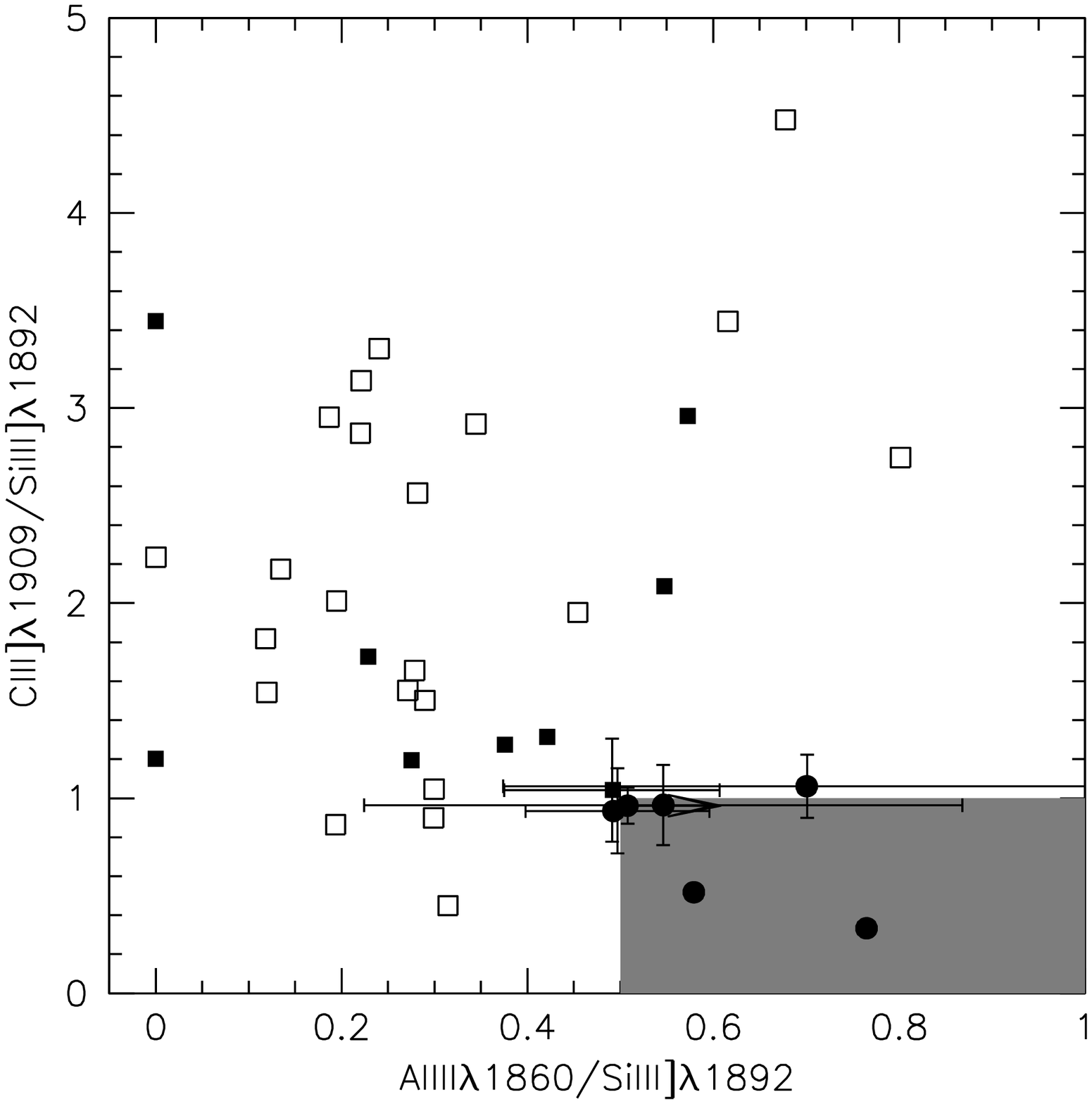}
\caption{{ The distribution of Pop. A sources (A1: open squares; A2: filled squares; A3: filled circles) in the plane defined by the ratios \ciii/\siiii\ and \aliii/\siiii. The area associated with xA sources is the lower-left shaded box. Error bars are shown for borderline sources only. The source with a lower limit to \aliii/\siiii\ is PG 1415+415, whose \aliii\ profile is affected by absorption lines. } \label{fig:test}}
\end{figure}

{ If optical and UV rest frame range are both covered, it  becomes possible to test that  the condition \rfe $\ga$ 1 is fully interchangeable with  \aliii/\siiii $\ga 0.5$ and \siiii/ \ciii\ $\ga 1$\ (i.e., \rfe $\ga$ 1  $\Leftrightarrow$   \aliii/ \siiii $\ga 0.5$ and  \siiii/ \ciii\ $\ga 1$). 
If only xA sources are considered, at low $z$ not many spectra covering the 1900 blend  are available in the MAST archive. At high $z$, few objects  have \hb\ covered in IR windows with adequate resolution spectroscopy. At present, data available to us encompass 6 low-$z$\ (including I Zw 1, and excluding all RL sources)  and 3 HE sources of sample 2. 

\paragraph*{Low-$z$ -- } { We measured with {\sc specfit} both UV ratios on $\approx$ 100 sources of \citet{bachevetal04} for which the 1900 blend is covered from  HST/FOS observations, as a test of consistency for the two criteria. Among them, six A3  sources (1H 0707--495, HE 0132--4313, I Zw 1, PG 1259+593, PG1415+451, PG1444+407) are included. Spectral type assignments and  references to the used data   can be retrieved from  \citet{sulenticetal07}.   

All Pop. B (56) sources save one\footnote{A misclassified source, SBS 0916+513, for which an A2 spectral type has been deduced from a new SDSS spectrum.}  show  \ciii/\siiii\ $\ga$2.0 and are therefore immediately excluded by the selection criteria; many sources appear dominated by \ciii\ emission, with \ciii/ \siiii\ $\gg$1.  Fig. \ref{fig:test} shows the distribution of 37 Pop. A sources   in the plane \ciii/\siiii\ vs \aliii/\siiii, with special attention to sources that are ``borderline'' and for which error bars are also shown. All type A3 objects fall  at the border or within the shaded region defined by the conditions  \aliii/\siii $\ge$0.5\ and \ciii/\siiii\ $\le$1. Sources falling close to the border show \rfe\ $\approx 1$, while the two sources with  \rfe $> 1$ are well within the region. The only source (Mark 478) that is not formally of spectral type A3 and falls close to the area  \aliii/\siii $\ge$0.5 \ciii/\siiii\ $\le$1 is of type A2  with \rfe$\approx$0.92.  Fig. \ref{fig:test} confirms the consistency of the optical and UV criteria: no source  of \rfe $<$ 0.9  falls in the domain \ciii/ \siiii$ \la$1 and \aliii/\siiii/ $\ge$ 0.5, and no source with \rfe$\ga$0.9 falls, within the error, outside of it. }

\paragraph*{High-$z$ --}  Three sources of sample 2 have newly-obtained observations covering the \civ -- \ciii\ emission lines (Mart\'{\i}nez-Carballo et al. 2014, in preparation): HE0122--3759 ($\log L \approx 47.7$ [\ergss]), HE0358--3959 ($\log L \approx 47.3$), HE1347-- 2457 ($\log L \approx 47.9$). In all cases, the condition on the UV lines is satisfied.  }

\medskip
{  The prototype source I Zw1 has $\log L \approx 45.6$ [\ergss]; the lowest luminosity and nearest  source is 1H 0707--495 with $\log L \approx 45.04$ at $z \approx 0.04$; the luminosity of the three PG  sources is the range $\log L \approx 45.5   - 47.0$. The condition \rfe $\ge$ 1 implied \aliii/\siiii\ $\ge 0.5$\ in all cases, and all individual spectra consistently showed the features typical of A3 sources (low \civ\ equivalent width, large \civ\ blueshift, etc.). Within the limit of the available data, the optical and UV criteria are consistent over a very large range in luminosity and redshift, $ 45 \la \log L \la 48$, $0.04 \la z \la 3$. }

%One could also select a much larger sample of lower redshift sources with \rfe$>$1.0 from the SDSS. Few of these sources are ideal from the points of view of S/N but a significant  sample of extreme candidates can be found there. Figure \ref{fig:lledd} shows a sample of sources with  $\log L_\mathrm{bol}>$46.0 taken from \citet{shenetal11}. The left panel shows the source distribution in the \rfe\ vs. \lledd\ plane while the right panels shows histograms of the \lledd\ distribution in intervals of \rfe\ -- the lower 2 represent bins A3$^\mathrm{m}$ and A4$^\mathrm{m}$ as defined by \rfe. Bins A3$^\mathrm{m}$\ and A4$^\mathrm{m}$\ certainly favor higher \lledd\ sources. The bin A4$^\mathrm{m}$\ sample represents the highest priority sample for further study.   Figure \ref{fig:lledd} serves only to illustrate that a significant number of extreme Eddington candidates can be identified and used in a preliminary tests of cosmological models. Increasing access to large aperture telescopes opens the possibility of verifying candidate samples and deriving much more accurate \rfe\ and \lledd\ (derived via FWHM H$\beta$ and \mbh) estimates. 
\subsection{The a-posteriori distribution of Eddington ratios}
\begin{figure}
%\epsscale{1.00}
\includegraphics[scale=0.45]{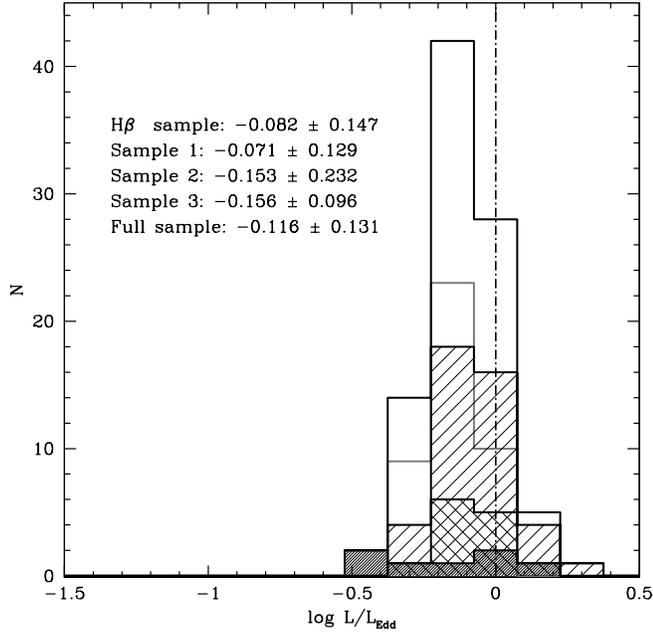}
\caption{Distribution of $\log$\lledd\ for the sources of the full sample considered in this paper. The  shaded histograms shows the distribution of \lledd\ for  subsamples  1  and 2 (dark shaded) based on \hb, { and sample 3 based on the 1900 blend (thick grey line)}.   The cross-hatched histogram represents the distribution of the A4$^\mathrm{m}$\ sources in sample 1 (sample 1a of Tab. \ref{tab:hb}).  \label{fig:lleddour}}
\end{figure}

%The scatter in the Figure \ref{fig:lledd} bin A3$^\mathrm{m}$\ and A4$^\mathrm{m}$\ histograms can certainly be reduced since A3$^\mathrm{m}$\ and A4$^\mathrm{m}$\ are affected by blueshifted emission of emitting gas (\S \ref{sample}).  
An estimate of the spread associated to the \lledd\ distribution comes from the a-posteriori analysis of our sample. We computed \mbh\  following \citet{assefetal11}\ and { applied the bolometric correction indicated by \citet{richardsetal06} (optical) and  derived using the \citet{mathewsferland87} continuum (UV, see discussion on $B. C.$\ later in this section). } The resulting \lledd\ distribution is shown in Fig. \ref{fig:lleddour}. 
%There are systematic differences between the samples that are related to the assumed $B. C.$\ between samples 1 and 2 (not significantly different) and sample 3. The difference is driven by sample 1 that has a somewhat higher average \lledd; a small increase by $\Delta \lambda_\mathrm{Edd} \approx$ 0.08 in the $B. C.$\ for sample 3 would render any difference not significant. 
The dispersion in the  full sample is $\sigma \approx 0.13$. This is consistent with current search techniques that allow us to identify sources radiating within $\pm$0.15 dex in  log \lledd\ \citep{steinhardtelvis10}.\footnote{ The dispersion of the \lledd\ distribution is not dependent on the cosmology assumed. Considering the cases of Table \ref{tbl-1} (and some of them are rather extreme and unrealistic), the dispersion of  the Eddington ratio distribution changes little, $\delta \sigma \approx \la 0.01$. As expected the average  of the \lledd\ distribution is instead significantly dependent on the cosmology assumed, with differences that are $\la 0.2$ dex.  }
  Inter-subsample systematic differences  are   smaller than the full sample dispersion.  A4 \hb\ sources of of sample 1 (sample 1a in Tab. \ref{tab:hb}; cross-hatched histogram in Fig. \ref{fig:lleddour}) do not differ significantly from A3 sources. There is a systematic difference between the \hb\ and \aliii\ sample, by $\Delta \log \lambda_\mathrm{Edd} \approx 0.07 < 1 \sigma$ \ that is dependent on the assumed ratio of the $B.C.$\ at 1800 \AA\ and 5100 \AA\ ($\approx$ 0.63). This can give to an important systematic effect discussed in \S \ref{system}.

Armed with selection criteria to isolate sources clustering around \lledd $\approx 1$\ and having defined preliminary samples, we now discuss how we can derive luminosity information without prior redshift knowledge.

\section{Measuring luminosity from emission line properties} \label{bozomath}
\label{lumv}

The bolometric luminosity -- black hole mass ratio  of a source radiating at 
Eddington ratio $\lambda_\mathrm{Edd}$\ can be expressed as:

\begin{equation}
\frac{L}{M} \approx 10^{4.53} \lambda_\mathrm{Edd} \left(\frac{L}{M_{\rm BH}}\right)_{\odot} \approx 
 10^{4.81} \lambda_\mathrm{Edd} ~\mathrm{erg ~ s}^{-1}~ {\mathrm g}^{-1}, \end{equation}

where $M_{\rm BH}$\ is the  black hole mass, and $L$\ the bolometric luminosity. Under the assumption 
of virial motion the bolometric luminosity is
(setting $\xi \approx 10^{4.81}$\ erg s$^{-1}$\ g$^{-1}$):

\begin{equation}
\label{eq:l}
L \approx \xi \lambda_\mathrm{Edd} M \approx \xi \lambda_\mathrm{Edd} f_\mathrm{S} \frac{r_\mathrm{BLR} (\delta v)^{2}}{G}. 
\end{equation}

where $f_\mathrm{S}$\ is the structure factor \citep{collinetal06}, $\delta v$ a virial velocity dispersion 
estimator, $G$\ is the gravitational constant, and $r_\mathrm{BLR}$\ the BLR radius. The ionization parameter can be written as (under the assumption -- satisfied by spherical symmetry --  that the line emitting gas is seeing the same continuum that we observe):

\begin{equation}
\label{eq:u}
U = \frac {\int_{\nu_0}^{+\infty}  \frac{L_\nu} {h\nu} d\nu} {4\pi n_\mathrm{H} c r_\mathrm{BLR}^2}
\end{equation}
where $L_{\nu}$\ is the specific luminosity per unit frequency, $h$\ is 
the Planck constant, $\nu_{0}$\ the Rydberg frequency, $c$ the speed of light, and \nh\ the hydrogen number density. The parameter $r_\mathrm{BLR}$\ can be interpreted as the distance between the central source of ionizing radiation and the part of the line emitting region that responds to continuum changes.  Values of $r_\mathrm{BLR}$\  from reverberation mapping \citep{petersonetal98} of \hb\ are available for $\approx$60 low-$z$\ Seyfert 1 galaxies and quasars  \citep{bentzetal09,bentzetal13}. The most recent $r_\mathrm{BLR}$\ determinations show a correlation with luminosity $r_\mathrm{BLR} \propto L^{0.533^{+0.035}_{-0.033}}$\ \citep{bentzetal13}, consistent with  $U$\ remaining constant with luminosity. 
%Our higher redshift study suggests an exponent closer to 0.6 (see Figure 11 in \citet{marzianietal09}.

There is an alternative way to derive $r_\mathrm{BLR}$ if one has a good estimate of the product of  
\nh\  $\times$$U$. Without loss of generality, 
\begin{equation}
\label{eq:r}
r_\mathrm{BLR} = \left[ \frac {\int_{\nu_0}^{+\infty}  \frac{L_\nu} {h\nu} d\nu} {4\pi Un_\mathrm{H} c} \right]^{1/2} =
\left( \frac {{\kappa L}} {4\pi Un_{\rm H} c h\bar{\nu_\mathrm{i}}} \right)^{1/2}
\end{equation}
where the ionizing luminosity is assumed to be $L_\mathrm{ion} = \kappa L$, with $\kappa \approx 0.5$. The number of ionizing photons is $\kappa L/h\bar{\nu_\mathrm{i}}$, where $\bar{\nu_\mathrm{i}}$\ is the average frequency of the ionizing photons. Several workers in the past used Eq. \ref{eq:r}  to estimate $r_\mathrm{BLR}$\  
\citep{padovanietal90,wandeletal99,negrete11}.  Analysis of a subsample of reverberation mapped sources 
indicates that Eq. \ref{eq:r} provides estimates of $r_\mathrm{BLR}$\ not significantly different from 
reverberation values \citep{negreteetal13}.  \citet{bochkarevgaskell09}\ also show that a photoionization analysis 
based on the \hb\ luminosity provides   $r_\mathrm{BLR}$\  estimates consistent  with reverberation mapping. 

Inserting Eq. \ref{eq:r} into Eq. \ref{eq:l}:

\begin{eqnarray}
L & \approx & \frac{\xi}{\sqrt{4\pi c h} G}\frac{\lambda_\mathrm{Edd} f_\mathrm{S}\kappa^{\frac{1}{2}}}{\bar{\nu_\mathrm{i}}^{\frac{1}{2}}
}  \frac{L^{\frac{1}{2}}}{(n_\mathrm{H}U)^{\frac{1}{2}}} (\delta v)^{2} \\ \nonumber
&\approx &  \frac{\xi^{2}}{4\pi c h G^{2}}\frac{\lambda_\mathrm{Edd}^{2} f_\mathrm{S}^{2} \kappa}{\bar{\nu_\mathrm{i}}}  \frac{1}{(n_\mathrm{H}U)} (\delta v)^{4}   
\end{eqnarray}

Then:

\begin{equation}
\label{eq:ln}
L \approx 7.8 ~10^{44} \frac{\lambda_\mathrm{Edd,1}^{2} \kappa_{0.5}  f^{2}_\mathrm{S,2}}{h {\bar\nu_{\mathrm{i}}, _\mathrm{100 \, eV}}}
\frac{1}{(n_\mathrm{H}U)_{10^{9.6}}} (\delta v)_{1000}^{4}~~ \mathrm{erg~ s^{-1}}
\end{equation}
where the energy value has been normalized to 100 eV ($\bar{\nu_\mathrm{i}} \approx 2.42 \, 10^{16}$\ Hz),  the product $(n_\mathrm{H}U)$ to the ``typical'' value $10^{9.6}$\ cm$^{-3}$\ \citep{padovanirafanelli88,matsuokaetal08,negreteetal12} and $\delta v$\ to 1000 km s$^{-1}$.  

 Eq. \ref{eq:ln} (hereafter the ``virial'' luminosity equation) is formally valid for any \lledd;  the key issue in the practical use of Eq. \ref{eq:ln} is to have a sample of sources tightly clustering around an average \lledd\ (whose value does not need to be 1, or to be {\em accurately }known). At present, we can identify sources with $\lambda_\mathrm{Edd} \rightarrow 1$, but it is still possible that an eventual analysis may employ different spectral types representative of much different \lledd\ average values.  In practice, an approach { followed in this paper has been} to consider Eq. \ref{eq:ln} in the form $L \approx \mathcal{L}_{0} \delta v^{4}$, where $\mathcal{L}_{0}$\ has been set by the best guess of the quasar parameters with $\lambda_\mathrm{Edd} \rightarrow 1$. This will imply a value of \h, and to ignore source-by-source diversity. A second possibility is to compute $L \approx \mathcal{L}_{0}$ to yield the concordance value of \h, especially for a sample of sources  at very low-$z$\  ($\ltsim 0.05$) where \om\ and \ol\ are insignificant in  luminosity computations. 
  Statistical and systematic errors will be  discussed in \S \ref{seleffects}.

%The luminosity of I Zw 1 can be retrieved from the Eq. \ref{eq:ln} considering \nh $U$\ $\approx$ 10$^{9.25}$ \cm3, FWHM $\approx$ 1050 \kms\ \citep{negreteetal12}, $\nu_\mathrm{i,2.42 ~10^{16}} \approx 0.43$\ (from the spectral energy distribution of \citealt{grupeetal10} appropriate for NLSy1s), and $\lambda_\mathrm{Edd} = 1$ we obtain $\log L \approx 45.59$\ which is consistent with 45.53 estimated from the observed flux, $z\approx0.050$, and concordance cosmology. 

\section{Preliminary Results}
\label{cosmo}

\subsection{Comparison of virial luminosity and luminosity derived
from redshift}

\begin{figure}
%\epsscale{.60}
\includegraphics[scale=0.45]{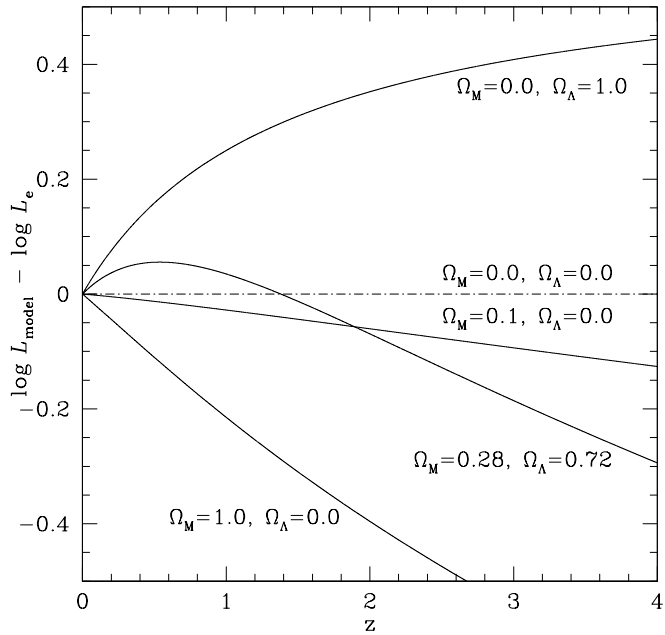}
\caption{Expected bolometric luminosity difference for several cosmological models, referenced to the case of an ``empty'' Universe (horizontal dot-dashed line) with  $\Omega_{M} = 0.00$, $\Omega_{\Lambda} =0.00$. \label{fig:diff}}
\end{figure}

Using our three combined samples we initially consider four cases: (1) 
``concordance cosmology''  (\om + \ol = 1.0, \om = 0.28,
\ok = 0.0), (2) an ``empirical'' open model  (\om = 0.1, \ol =
0.0, \ok=0.0), (3) a matter-dominated Universe  (\om = 1.00,
\ol = 0.0, \ok=0.0) and (4) a $\Lambda$-dominated model \om =
0.0, \ol = 1.0, \ok=0.0). Expected bolometric luminosity differences as a function of redshift  with
respect to an empty Universe are shown in Figure \ref{fig:diff}.

The transverse comoving distance can be written as
\begin{equation}\label{eq:dp}
d_\mathrm{p} = \frac{c}{H_{0}} f(z, \Omega_{\Lambda}, \Omega_\mathrm{M}),
\end{equation}
where $f(z,  \Omega_{\Lambda}, \Omega_\mathrm{M})$\ is a function of $z$\ with $ \Omega_{\Lambda}$\ and 
$\Omega_\mathrm{M}$ assumed as parameters reported in
\citet{perlmutteretal97}.  Bolometric luminosity is defined by the relation:
\begin{equation}\label{eq:zlum}
\log L = \log\left[4 \pi \lambda f_{\lambda} d_\mathrm{p}^{2} (H_0, z, 
\Omega_\mathrm{M}, \Omega_{\Lambda})\right] + B.C.,
\end{equation}
where $B.C.$\ is the bolometric correction and $f_{\lambda}$\ the {\em
rest frame} specific flux at 5100 \AA\ or 1800 \AA. Assumed bolometric corrections are: $B.C.$ = 1.00  for $\lambda f_{\lambda}$\ measured at 5100 \AA\ { \citep{richardsetal06}, and $B.C.$ =  0.800 when $\lambda f_{\lambda}$\ was measured  at 1800 \AA. These bolometric correction are computed for the \citet{mathewsferland87} continuum that is believed appropriate for Pop. A quasars.}

\begin{figure}
%\epsscale{.60}
\includegraphics[scale=0.45]{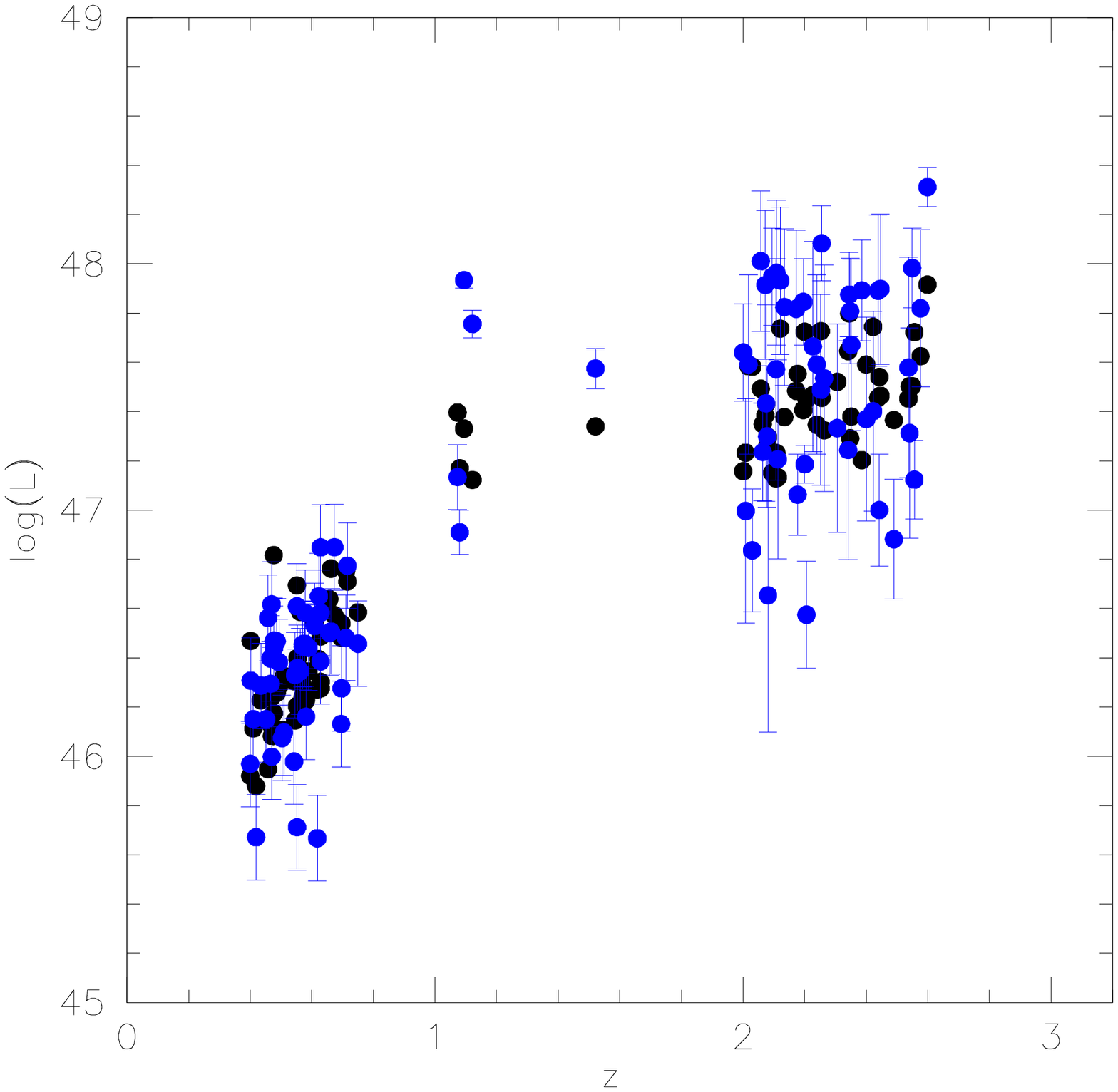}\\
\includegraphics[scale=0.45]{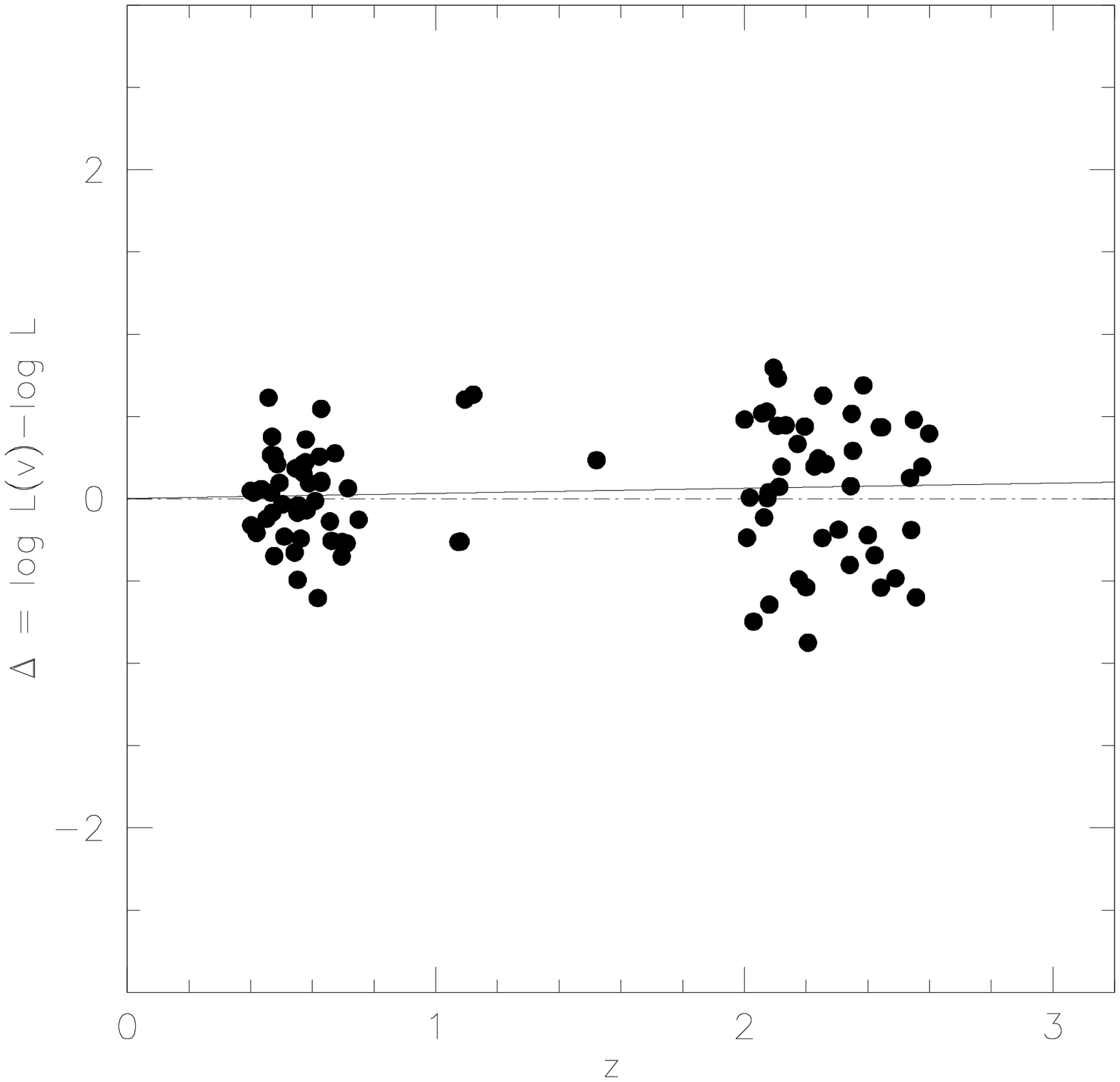}
\caption{Bolometric luminosity (in logarithmic scale) for our sample of 92 quasars, computed from Eq. \ref{eq:ln} (blue circles) and from the customary relationship using concordance cosmology (assuming $H_{0}$ = 70 \kms Mpc$^{{-1}}$). The bottom panel shows the residuals. The line is an (unweighted) linear least-square (lsq) fit to the residuals. \label{fig:highal}}
\end{figure}

We compare two sets of luminosity values as a function of redshift: one
derived from the redshift $L$\ (Eq. \ref{eq:zlum}) and the other from the
virial luminosity (Eq. \ref{eq:ln}). We assume ${\cal L}_0 \approx 1.16 \cdot 10^{45}$ \ergss\ from the best guess of parameters entering Eq.  \ref{eq:ln}, for the standard continuum of \citet{mathewsferland87}: $\kappa \approx 0.6$, \nh $U$\ $\approx 10^{9.6}$ \cm3, $h\bar{\nu_\mathrm{i}} \approx 41$ eV, and $f_\mathrm{S}\approx 1.5$ as recommended by \citet{collinetal06}.  Figure \ref{fig:highal} shows data points 
computed from the virial equation (blue circles)  and values computed from the redshift in the case of
concordance cosmology ($H_{0}$=70 \kmpc, \om=0.28, \ol=0.72). Error bars are 1$\sigma$\
uncertainties from FWHM uncertainty estimates given in \S
\ref{sample} and are reported in Tab. \ref{tab:hb} and \ref{tab:aliii} { (Fig. \ref{fig:highal})}. Residuals
defined as $\Delta = \log L(v) - \log L$\ are shown in the bottom panel.

The average of residuals ($\bar{\Delta}$) is nonzero and there is a
slightly nonzero slope   $ b \approx    +0.03$ in the lower panel of Fig. \ref{fig:highal}, { 
not significantly different from 0\ since its 1$\sigma$\ error is $\sigma_\mathrm{b} \approx  $0.075}.  
The shape of the residuals depends on
redshift, as it is related to the shape of $f(z, \Omega_\mathrm{M},
\Omega_{\Lambda})$\ and
hence to the metric and  the $\Omega$s:

\begin{eqnarray}
\Delta (z, H_0, \Omega_\mathrm{M}, \Omega_{\Lambda})  & =  & \log L(v) - [ \log(4 \pi c^2) + \log(H_0^2)  \nonumber \\ 
&  & + \log(\lambda f_\lambda) +\, B.C. + \log f^2( z,  \Omega_\mathrm{M}, \Omega_{\Lambda})] \nonumber \\ 
& = & const. - \log(H_0^2)   - \log f^2( z,  \Omega_\mathrm{M}, \Omega_{\Lambda})
\label{eq:delta}
\end{eqnarray}

The above equation shows that  $H_0$\ sets the scale  for $\Delta \log L$\ in a way
that is not dependent on redshift.  A change in the model involving different $\Omega$\ values leaves
correlated residuals:
\begin{equation}
\Delta(z, H_0, \Omega_\mathrm{M}, \Omega_{\Lambda}) = \bar{\Delta}(H_0, \Omega_\mathrm{M}, \Omega_{\Lambda}) + \zeta(z, \Omega_\mathrm{M}, \Omega_{\Lambda}). 
\end{equation}
$\bar{\Delta}$   is not formally dependent on $z$ since it is defined as an average, and its value is fixed once the redshift distribution of a sample is given.  Retrieving $\zeta(z, \Omega_\mathrm{M}, \Omega_{\Lambda})$\ when $\Delta(z, H_0, \Omega_\mathrm{M}, \Omega_{\Lambda}) \equiv 0$\ is equivalent to finding \om\ and \ol\ and hence
the appropriate model. { More explicitly,   the difference $\zeta(z, \Omega_\mathrm{M}, \Omega_{\Lambda})$ between a cosmology independent luminosity estimator (the virial luminosity) and a luminosity computed from luminosity distance should become identically zero as a function of redshift for the correct cosmological model. }

{  We consider  the conventional $\chi^{2}$\ computed from the squared $\Delta$s  as a goodness of fit estimator. In principle the residuals $\Delta(z, H_0, \Omega_\mathrm{M}, \Omega_{\Lambda})$\ could be fit with an analytical approximation of $\zeta(z, \Omega_\mathrm{M}, \Omega_{\Lambda})$\ as complex as needed.}  Our preliminary  data   allow  { only a simple linear least square fit  to the residuals:
$\Delta(z) = a + b \cdot z $. The sought-for model requires $a = b =0$.}
In the following analysis we found   expedient to consider the normalized average $\bar{\Delta}/\sigma_{\bar{\Delta}}$\  and the normalized slope $b/\sigma_\mathrm{b}$.    $\bar{\Delta}/\sigma_{\bar{\Delta}}$\ and $b/\sigma_\mathrm{b}$\ are both  $t$-distributed estimators that for our sample size can be considered normally distributed. They both  provide statistical confidence limits  $\propto 1/({\mathrm{rms}}/\sqrt{N-1})$, where $N = 92$.  The normalized average and the slope estimator have  been computed under the assumption that the rms scatter of our data is an estimate of uncertainty for individual measures \citep[][Ch. 15]{pressetal92}. { The slope estimator is not a { very} tightly constraining parameter; however it has the considerable advantage to be fully independent on \h, and to provide a straightforward representation of the systematic trends associated with \om\ and \ol\ as function of $z$. }
%and yield  tighter confidence limits than the normalized $\chi^{2}$. 
%is the total number of quasars in the  three samples. 

The rms value ($\approx 0.365$)  is intrinsic to our dataset and will probably not change
much with larger samples unless FWHM measures with substantially higher accuracy, 
and a reduction of other sources of statistical errors, are obtained (\S \ref{seleffects}). 
Given the redshift range of our data, a change in \om\ gives rise to a change in slope of the residuals $\Delta(z)$\ that strongly affects $\bar{\Delta}$\ (as can be  glimpsed from Fig. \ref{fig:diff}). 
%\label{eq:linear}

%(although in \S \ref{disc} we mention how in principle, the virial luminosity could be made independent from any assumption on \h). } 
%In principle  weights should be applied to data in the least square fit, however, the statistical bias of our sample (the strongly uneven $z$\ distribution) and the higher accuracy of sample (2) lead to implausible results using a weighted fit. We  prefer to use the rms scatter of our data as an estimate of uncertainty for the confidence limit computation (following the prescription of \citealt{pressetal92}, Ch. 15) and use an unweighted least square fit. Restriction to sample (3b) yields consistent results with sample (3a) and no major advantages due to the smaller (3b) sample size, so that we will only consider the larger (3a) sample in the following. 

 %for the case of a flat Universe, \om+\ol = 1 (\S \ref{flat}), and for unconstrained determination of \om\ and \ol\  (\S \ref{omol}).

\subsection{Selected alternative cosmologies}
\label{selected}

%with three different $H_{0}$\ values. % The 
%mean values have been normalized by the standard deviation,
%$\sigma_{\bar{\Delta}}$.  %The first column provides the slope $b$\
%of the linear fit (Eq. \ref{eq:linear})  that only depends on \om\ and \ol.

Table \ref{tbl-1} reports   normalized average $\bar{\Delta}/\sigma_{\bar{\Delta}}$, $b$, and $\chi^{2}$   values for the selected cosmological models  of Fig. \ref{fig:diff}, assuming fixed \h = 70 \kmpc.  Normalized  $\chi^{2}/\chi^{2}_\mathrm{min}$ values of 1.1, 1.41, and 1.79 limit acceptable fits at confidence levels of 1$\sigma$, $2\sigma$, and  $3\sigma$ respectively.   Note that  $\chi^2$\ and $\bar{\Delta}/\sigma_{\bar{\Delta}}$ values are affected by  \h, while $b$\ is not dependent on \h.  

Extreme implausible cases such as a matter (\om=1) or $\Lambda$-dominated (\ol=1) Universe are ruled out.  The case with \om = 1 (\ol= 1) will yield a large positive (negative) slope meaning that redshift based estimates are underluminous (over luminous) with respect to virial luminosities.   The concordance case is favored by our dataset, with $\chi^2 \approx 1.05$\ (Fig. \ref{fig:highal}). The ``empirical'' model with $\Lambda= 0$\ and \om = 0.1 is not disfavored by our data, with $\chi^2 \approx 1.08$ if \h = 70 \kmpc.  The two cases, concordance and ``empirical'' \om = 0.1 { (along with the unrealistic case of a fully empty Universe)} cannot be statistically distinguished since  their slopes $b$\ are close to 0 and $|b/\sigma_\mathrm{b}| \la 1$. 

{ Obtaining a solution close to the one of concordance cosmology is not a consequence of how the parameters entering Eq. \ref{eq:ln} were obtained. Parameters related to continuum shape and ionising photon flux were not derived assuming any particular luminosity. The assumed value of ${\cal L}_0$ depends  from \h\ since   the assumed \lledd\    depends on \h.  However,  effects of \om\ and \ol\ are negligible since    \lledd\ was calibrated on a very low-$z$ source, I Zw 1  (Fig. \ref{fig:diff}). }
 
\subsection{$H_{0}$, \om\ and \ol\ in a flat Universe}
\label{flat}

\begin{figure}
%\epsscale{.75}
%\includegraphics[scale=0.4]{f3a.eps}
%\includegraphics[scale=0.4]{f3b.eps}\\
\includegraphics[scale=0.31, angle=0]{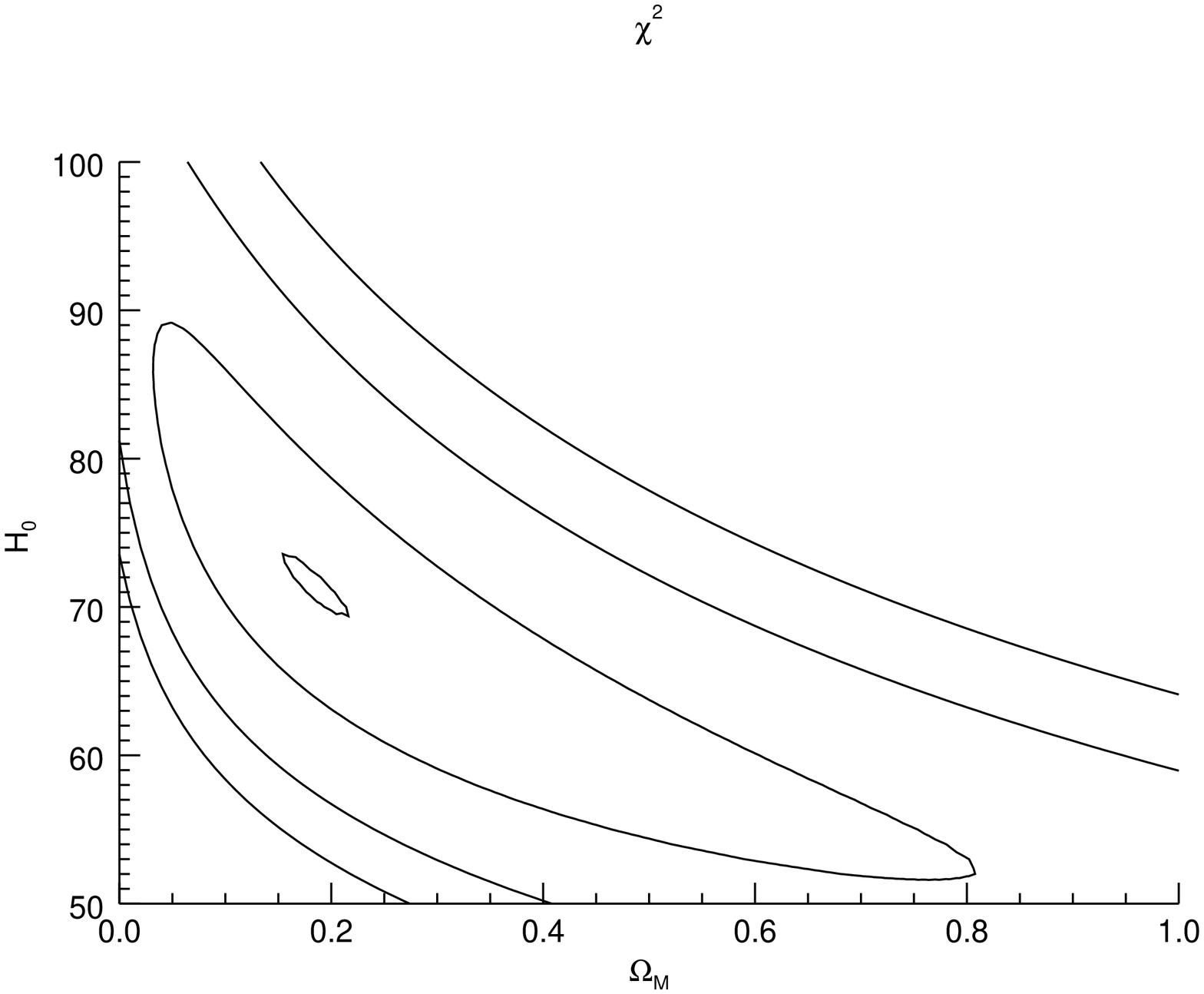}\\
\includegraphics[scale=0.31, angle=0]{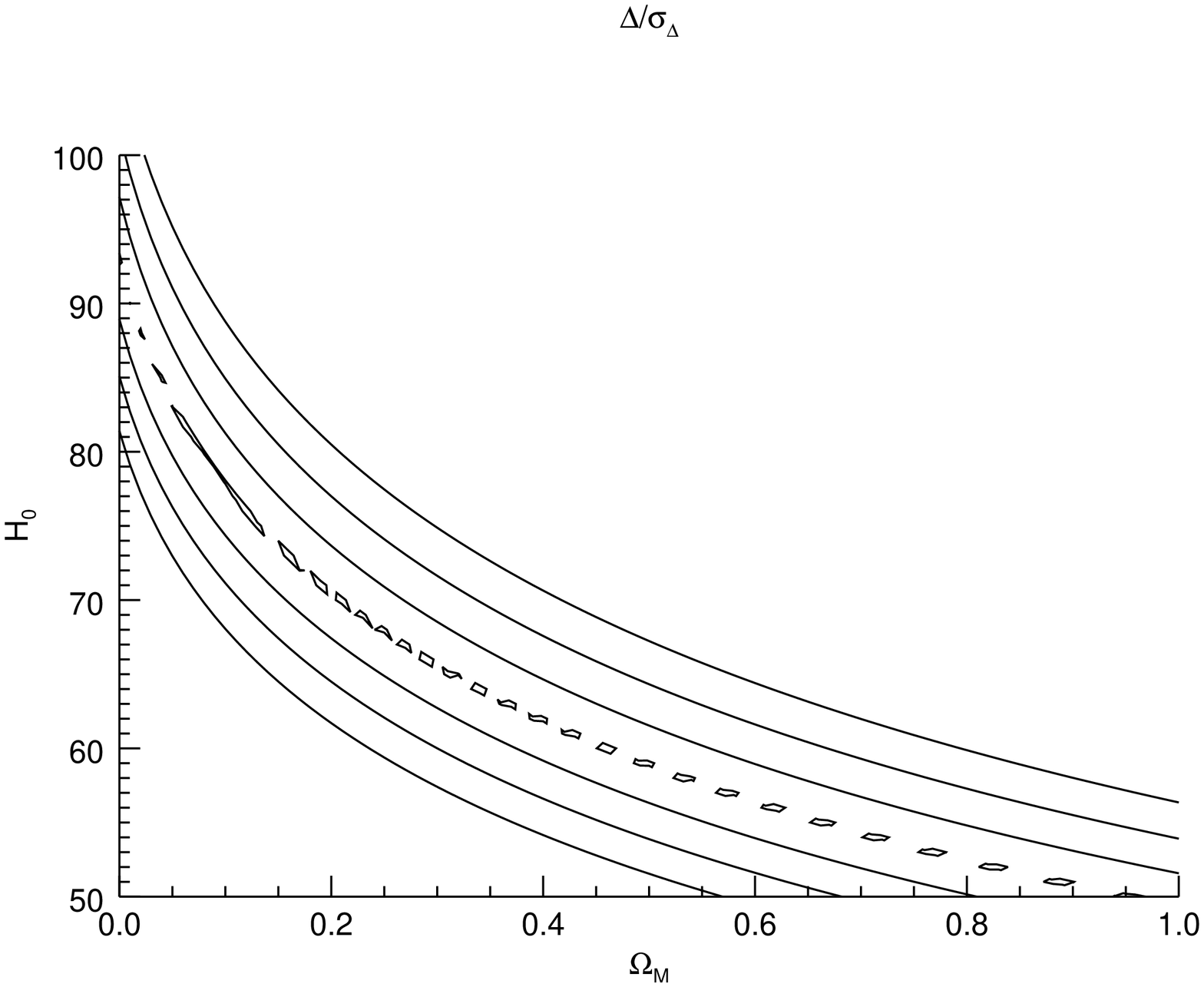}\\
\includegraphics[scale=0.35, angle=0]{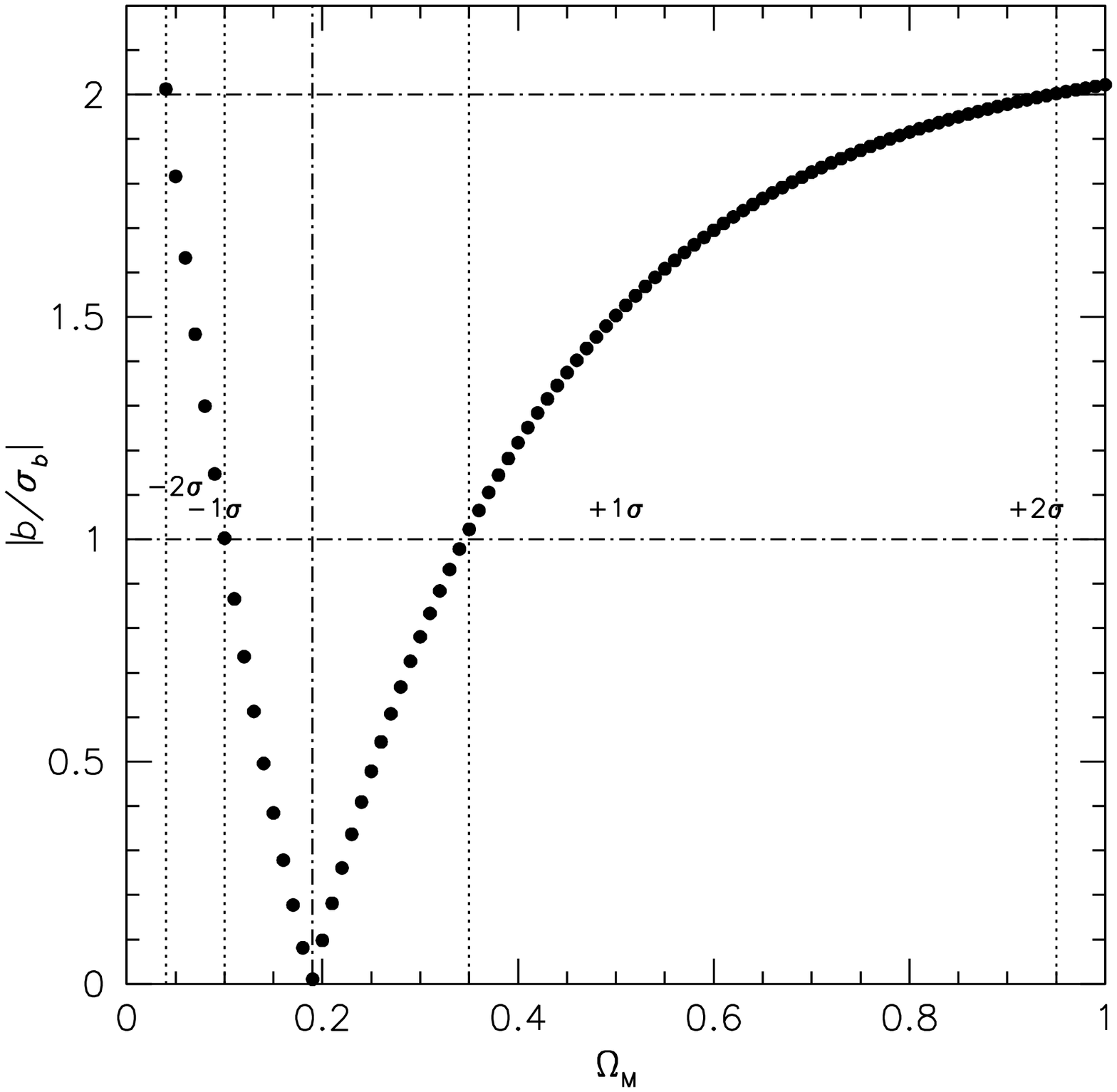}

\caption{Top: Normalized $\chi^2$  of luminosity residuals computed between virial estimates and models, for a flat geometry ($\Omega_\mathrm{M} + \Omega_{\Lambda} =1.00$) as a function of $H_{0}$ and $\Omega_\mathrm{M}$. 
Middle: same for average of residuals normalized to standard deviation. One contour line brackets the most likely value and the other three trace the uncertainties at 1$\sigma$, 2$\sigma$ and $3 \sigma$ confidence levels.
{ Bottom: Behavior of normalized slope parameter  $b/\sigma_\mathrm{b}$\ as a function of \om. The lower --2, --1$\sigma$\ and $+1, +2\sigma$\ limits are shown by dotted lines. }
%The  lines in the top  panels trace constant\ $\bar{\Delta}/\sigma_{\bar{\Delta}}$ $\approx$ at 0.1, 1, 2, 3. 
%Bottom: joint probability $P(\bar{\Delta}>0,b>0)$ = $P(|\bar{\Delta}|>0) \cdot P(|b|>0)$ \ for the $t$-distributed parameters $b/\sigma_\mathrm{b}$\ and normalized average. 
%$\chi^{2}$\ as a function of \om\ and \ol. The lines trace the $\chi^{2}/\chi^{2}_\mathrm{min} \approx$ 1.01, 1.13, 1.54, 2.03. The latter three value corresponds to confidence limits of 1,2,3 $\sigma$\ respectively.  
\label{fig:delta}}
\end{figure}

Assuming a flat geometry (\om + \ol = 1, \ok = 0) the normalized average
is influenced by changes in $H_{0}$ and \om\ (\ol\ is set by \om). 
  Fig. \ref{fig:delta}  shows $\chi^2$\  (top),   $\bar{\Delta}/\sigma_{\bar{\Delta}}$\ (middle) as a function of \h\ and \om, and $|b/\sigma_\mathrm{b}|$\ as a function of \om.  The $\chi^2$\  contour lines trace 4 values of the ratio $\chi^{2}/\chi^{2}_\mathrm{min}$: $\chi^{2} \approx
\chi^{2}_\mathrm{min}$, and 3 values meant to represent the 0.32, 0.05, and 0.003
probability interval considering that the probability of the
$\chi^{2}/\chi^{2}_\mathrm{min}$\ ratio follows an F-distribution
\citep{bevington69}.  { The $\chi^2$ diagram indicate that \om\ is constrained within 0.05 and 0.8 if \h\ is not chosen a priori.} The lines in the  $\bar{\Delta}/\sigma_{\bar{\Delta}}$\  plot  { (middle  panel of Fig. \ref{fig:delta})} identify a curved strip close to 0, and values 1, 2, 3. The lines of value 1,2,3 trace the corresponding 1,2,3 $\sigma$ confidence levels. The bottom panel  of Fig. \ref{fig:delta} shows the behaviour of  $|b/\sigma_\mathrm{b}|$ as a function of \om. { The minimum identifies the condition $b=0$. Changing \om\ will increase the slope that will become significantly different from 0 at 1 and 2 $\sigma$\ confidence levels when  $|b/\sigma_\mathrm{b}|$ =1 and 2. It is possible to define only --2, --1, and $ +1$, $+2\sigma$ uncertainties: \om $\approx$ 0.19$^{+0.16}_{-0.08}$, and  $\approx$ 0.19$^{+0.76}_{-0.15}$ at 1 and $2\sigma$, respectively.    }

Once an \h\ value is assumed as indicated by $\chi^{2}_\mathrm{min}$  or $ b = 0$, uncertainties on \om\  are significantly reduced  if the normalized average is used as a confidence limit estimator. Eq. \ref{eq:ln} in the form $L =  {\cal L}_{0} \delta v^{4}_{1000}$, with ${\cal L}_{0} \approx 1.16 \cdot 10^{45}$ \ergss,  implies { b = 0} and a minimum $\chi^2$\ at \h\ $\approx$ 71 \kmpc, yielding \om $\approx$ 0.19$^{+0.07}_{-0.05}$ at 1$\sigma$, and 0.19$^{+0.21}_{-0.12}$\ at 3$\sigma$\ (middle panel of Fig. \ref{fig:delta}).  Broader limits are inferred from the normalized $\chi^2$\ once \h\ is prefixed 
%(0.19$^{+0.14}_{-0.11}$\ and 0.19$^{+0.39}_{-0.18}$\ at 1 and 3$\sigma$) 
because $\chi^2$\ includes the uncertainty in \h\ that we may instead assume a priori.  Uncertainties from  $\chi^2$, $|b/\sigma_\mathrm{b}|$   and    $\bar{\Delta}/\sigma_{\bar{\Delta}}$\   reflect statistical errors only.  Constraints are already significant although uncertainties are not as low as those derived from the combination of recent surveys \citep{hinshawetal12,planck13}.

\subsection{Constraints on \om\ and \ol}
\label{omol}

Rather loose constraints are also  obtained where  \om\ and \ol\ are
left free to vary once   \h\ is assumed { (\h = 70 \kmpc)}.  The top panel of shows the $\chi^{2}$\
behavior in the plane \om\ and \ol.  Present data yield \om $\approx$ 0.19$^{+0.24}_{-0.19}$\
and  \om $\approx$ 0.19$^{+1.04}_{-0.19}$ at 1 and 3$\sigma$\ confidence level. {\ol\ is unconstrained within the limits of Fig. \ref{fig:omegas}. }

\section{Prospects for improvement: analysis of synthetic datasets}
\label{mock}

\begin{figure}
%\epsscale{.75}
%\includegraphics[scale=0.3]{mocksample200_0.2.eps}
%\includegraphics[scale=0.53,trim=100 2000 100  2000 ,clip]{h0ommedia200_0.2.eps}\\
%\includegraphics[scale=0.3]{slope200_0.2.eps}\\
%\includegraphics[scale=0.4]{h0om100_0.1.eps}
%\includegraphics[scale=0.3]{chifree200_0.2.eps}\\
\hspace{0cm}\includegraphics[scale=0.45,angle=0]{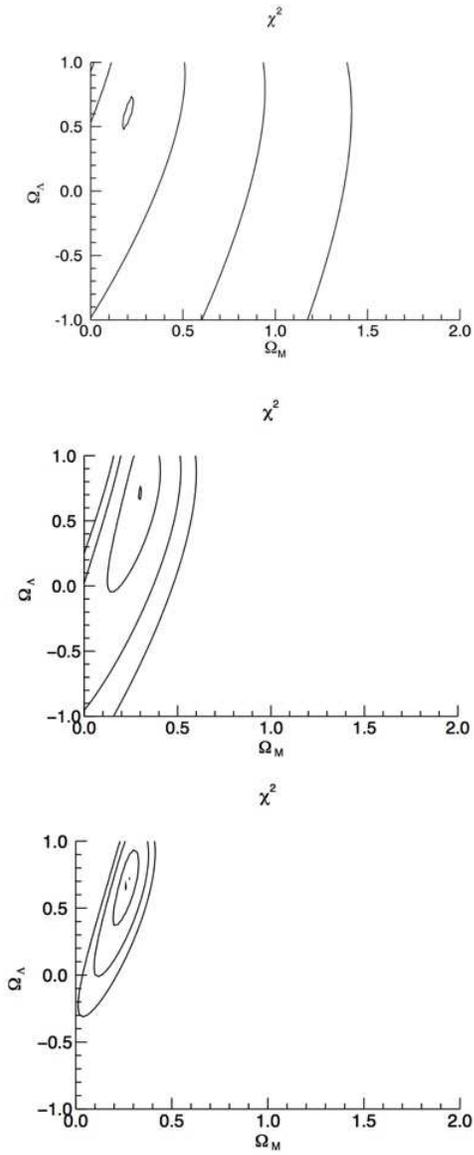}
\caption{Normalized $\chi^{2}$ behavior in the  \om\ -- \ol\ plane for data sample presented in this work (top) and mock samples with rms=0.2 (middle) and rms=0.1 (bottom).  Contour lines trace $\chi^{2}/\chi^{2}_\mathrm{min}$\ at confidence limits  1,2,3 $\sigma$.   \label{fig:omegas}}
\end{figure}

The limits on   \om, and \ol\ in Fig.  \ref{fig:omegas} are not strong
compared to previous studies despite the fact that \ol\ and \om\ are
close to currently accepted ones. Our preliminary sample  is too small and inhomogeneous. 
Extreme Eddington sources are estimated to include a sizable minority of the quasar population -- most likely $\approx$10\%. Significant improvements can come  
from an increase in sample size with  high S/N
observations, and by reducing 
the (large) statistical error in the dataset
analyzed here (\S \ref{budget}). 

To quantify the improvement that can be expected from the reduction in statistical errors, synthetic data were created by adding Gaussian deviates  to the  luminosity computed for the concordance case (\h = 70, \kmpc,  \om $= 0.28$,  \ol $= 0.72$), assuming a uniform redshift distribution.\footnote{A K-S test comparing the distribution of residuals    in our actual  sample and a Gaussian distribution indicates that the two are not statistically distinguishable. The same conclusion is reached through a $\chi^{2}$ analysis.} {  In other words, artificial  luminosity differences were made to scatter with a Gaussian distribution around  the luminosities derived assuming concordance cosmology.} This approach is meant to model sources of statistical error that may be present in the virial and $z$-based computations (i.e., FWHM uncertainty, errors in
spectrophotometry and bolometric luminosity). 

{ Figure \ref{fig:mock200} (top left panel) shows the residual as a function of redshift and reports average and dispersion for a mock sample of 200 sources with rms=0.2.  } The  { top right, lower left, and  lower right panels of Figure \ref{fig:mock200} show the behaviours of $\chi^2$,  $|b/\sigma_\mathrm{b}|$, and  $\bar{\Delta}/\sigma_{\bar{\Delta}}$  respectively, computed assuming \om+\ol=1}. 
Meaningful limits can be set in this hypothetical  case. { The $|b/\sigma_\mathrm{b}|$\ estimator yields \om $\approx 0.30^{+0.20}_{-0.12}$ at a 2$\sigma$ confidence level, with no assumption on \h.} If \h\ = 70 \kmpc, $\bar{\Delta}/\sigma_{\bar{\Delta}}$ (top right panel of  Fig. \ref{fig:mock200} )
indicates  \om $\approx 0.28^{+0.02}_{-0.02}$\ (1 $\sigma$).  

The middle panel of Fig. \ref{fig:omegas}, computed for unconstrained \om\ + \ol,
indicates \om $\approx 0.30^{+0.12}_{-0.09}$\ at 1$\sigma$\ confidence level, with  poor 
constraints on  \ol.  { The  $\chi^2$ \ distribution for the mock samples was computed as for the real data, varying \om\ and \ol\ with no constraints on their sum.}   %The lower right  panel  maps the joint probability  as a
%function of \om\ and $H_{0}$ similar to Fig. \ref{fig:delta}.  

The upper left panel of Fig. \ref{fig:mock100}\ (organised as  Fig. \ref{fig:mock200} for the \om + \ol = 1 case) shows a mock sample of 100 sources with  rms$\approx$0.1  { that is otherwise identical to the previous case}.  { The $|b/\sigma_\mathrm{b}|$\ estimator yields \om $\approx 0.28^{+0.13}_{-0.09}$ at a 2$\sigma$ confidence level, with no assumption on \h. If \h\ = 70 \kmpc, $\bar{\Delta}/\sigma_{\bar{\Delta}}$  (top right panel of  Fig. \ref{fig:mock100})  yields \om\
with  an uncertainty $\pm 0.05$ at a 3$\sigma$\ confidence level. }
From the $\chi^2$ distribution (lower right panel of  Fig. \ref{fig:mock100}), we derive \om$ \approx
0.28^{+0.07}_{-0.06}$, and \om$ \approx 0.28^{+0.005}_{-0.015}$\ if 
\h\ = 70 \kmpc\ and \om+\ol=1.   { The bottom panel of Fig. \ref{fig:omegas}  (\om + \ol\ unconstrained) indicates \om $\approx 0.28^{+0.04}_{-0.08}$\ at 1$\sigma$\ confidence level, again  with  poor 
constraints on  \ol. }
%\footnote{The approach of fitting a straight line to
%the residuals can be pursued also if rms$\approx$0.1 but Fig. \ref
%{fig:diff} indicates that a linear function will become a biased estimator
%if rms$\ltsim$0.1.}  % 

The mock sample improvement over our real sample comes from the larger sample size $N$ (a factor
$\sqrt{\frac{N}{92}}$), lower intrinsic rms (a factor 2 -- 4), and the
uniform distribution in redshift.  A reduction of errors associated  with the method is therefore tied to an increase in sample size since   uncertainties of $\bar{\Delta}$\  and  $|b/\sigma_\mathrm{b}|$\ decrease with $1/\sqrt{N}$.  Achieving  rms$\approx$0.1  would  require careful consideration of several systematic effects summarized below, or the definition of a large sample ($\sim$ 1000) of quasars (\S \ref{budget}).  

The value of   \ol\  is affected by   larger uncertainty than \om, but significant constraints (for example, the verification that \ol$> $0 at a 2$\sigma$\ confidence level) are within the reach. 
%of a hypothetical 400 quasars sample with rms$\approx$0.2.  
The weak constraints on \ol\ stem from 1) the large errors, 2)
the redshift range. Mock sample sources are assumed to spread uniformly over the $z$\ range 
0.1 -- 3.0, but only $z \ltsim 1.2$\ is  strongly affected by \ol\ (Fig. \ref{fig:diff}) so that only $\frac{1}{3}$\ of the sources are in the relevant range. 

We will set more stringent limits on the requirements for an 
observational program after discussing sources of statistical and systematic uncertainty.

\section{Limitations and sources of error/uncertainty}
\label{seleffects}

%We now make an overview of sources of error that will affect the use of extreme 
%quasars as standard candles. 
%A more quantitative analysis will be made when more definitive samples
%of extreme quasars have been identified. 
%Estimation of the $\Omega$s is affected 
%by statistical errors that will be significantly reduced with better sample selection and better source spectra. More explicitly,  the criteria used for selecting  extreme Eddington  sources  require that several parameters  in Eq. \ref{eq:ln} are fixed: $\kappa$ (\nh$U$), and $\bar{\nu_\mathrm{i}}$ --  changing them will imply a 
%change in line ratios. Fixing the value of  $\lambda_\mathrm{Edd}$\ most  likely fixes  $f_\mathrm{S}$\ \citep{netzermarziani10}. Eq. \ref{eq:ln} is therefore probably not influenced by systematic errors except for an offset  associated with the assumed 
%value of $\lambda_\mathrm{Edd}$ and this offset will only affect the estimation of $H_{0}$. Eq. \ref{eq:ln} is expected to provide a valid redshift-independent luminosity estimator suited for measurement of  \om\ and \ol. 

\subsection{Statistical error budget}
\label{budget}

Estimation of the $\Omega$s is affected 
by statistical errors that will be significantly reduced with larger samples   and better source spectra. The residual distribution is affected by errors on the parameters entering into the virial equation (Eq. \ref{eq:ln}) and into the customary determination of the quasar luminosity from redshift (Eq. \ref{eq:zlum}). In the following each source of statistical error is discussed separately, first for the virial equation and then for the $z$-based $L$\ determination. Col. 2 of Table \ref{tab:budget}\ reports the 1$\sigma$\ errors as estimated in the following paragraphs, propagating them quadratically to obtain an estimate of  statistical error that should be compared with the rms derived from our sample. 

\subsubsection{Virial equation}

\paragraph{Eddington ratio $\lambda_\mathrm{Edd}$}  -- The \lledd\ value of I Zw 1 is consistent with unity  (\S \ref{sample}), although the exact value depends on the normalization assumed for \mbh\ and on the bolometric correction. %It is unlikely that $\lambda_\mathrm{Edd} \gg 1$\ since a more refined estimate will increase \mbh\ leaving the bolometric  luminosity unchanged. %Assuming the derived $\lambda_\mathrm{Edd} \approx 0.8$\ for I Zw 1 as typical (a similar value is 
%obtained for the median of A4 sources in sample (a)), $\Delta  \approx -0.22$\ with respect to the assumed  $\lambda_\mathrm{Edd} \approx 1$.  
The \lledd\ distribution of Fig. \ref{fig:lleddour} shows $\sigma \approx$ 0.13. This dispersion value includes orientation effects.  It is unlikely that a larger sample can be obtained with a significantly lower scatter unless scatter is reduced if part of the dispersion can be accounted for by systematic trends (\S \ref{system}). 

\begin{figure}
%\epsscale{.60}
\includegraphics[scale=0.65]{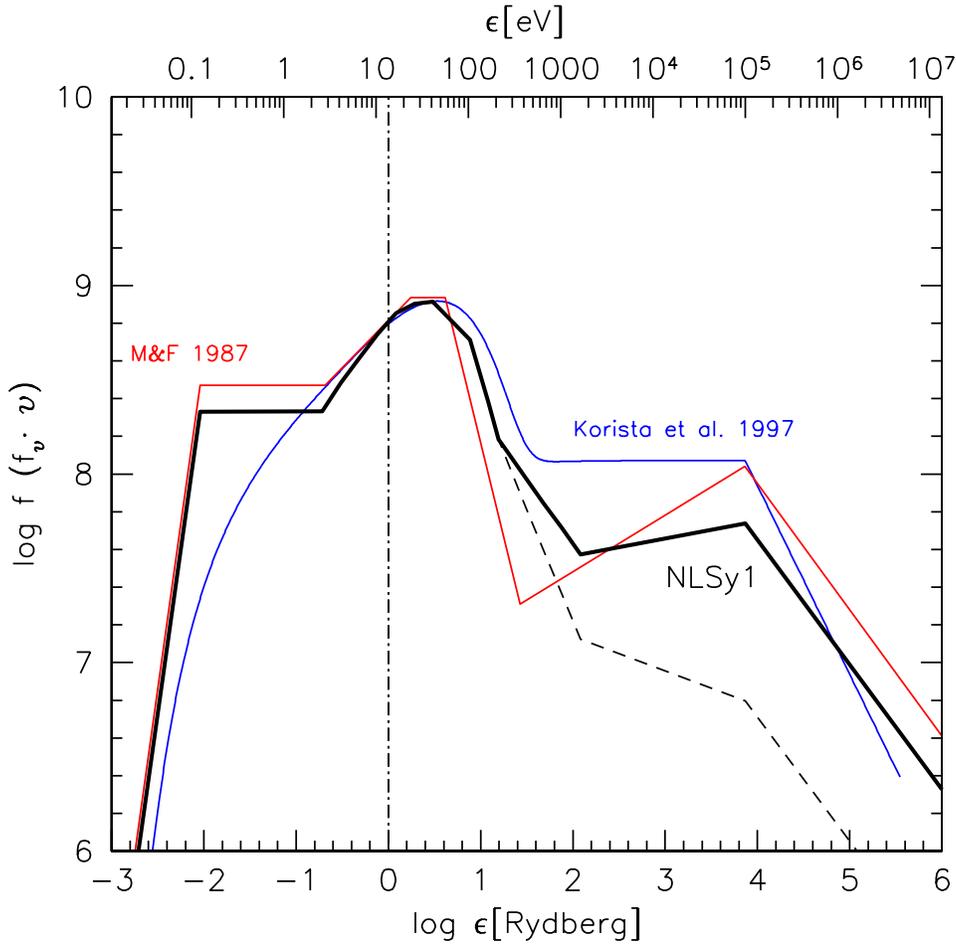}
\caption{Continuum shapes employed for the evaluation of the error budget, with   emitted power in arbitrary units as a function of photon energy in Ryd (lower $x$-axis) and in keV (upper $x$-axis). The thick solid line (NLSy1) is a SED appropriate for highly-accreting quasars; see text for a description of its construction. { The dashed line traces the steepest continuum  due to source by source diversity (i.e.,   the continuum  of the most extreme Pop. A sources). } The red and blue lines trace the    \citet{mathewsferland87} and the \citet{koristaetal97}\ continua, respectively.    \label{fig:contc}}
\end{figure}

\paragraph{Factor $\kappa/\bar{\nu_{\rm i}}$} -- The product $ U$\nh\ value entered in Eq. \ref{eq:ln} comes from   dedicated sets of photoionization simulations that assume \citet{mathewsferland87} continuum which is thought to be appropriate for Pop. A sources. However, the values of $\kappa$, $\bar{\nu_\mathrm{i}}$, and hence of  factor $\kappa/\bar{\nu_{\rm i}}$\ depend on the shape of the assumed photoionizing continuum.  Extreme sources are expected to converge toward large values of  $\Gamma_\mathrm{soft}$\ \citep[e.g.,][]{bolleretal96,wangetal96,grupeetal99,sulenticetal00a,grupeetal10,aietal11}.  In order to investigate the effect of different continua we considered a typical NLSy1 continuum as proposed by \citet[][Fig. \ref{fig:contc}]{grupeetal10}.  The average $\Gamma_\mathrm{soft}$\ derived by \citet{panessaetal11}\ was assumed for energies larger than 20 keV.   The dashed line  shows  a second ``minimum'' continuum representative of sources with $\Gamma_\mathrm{hard} \approx$ 2.2, computed on the basis of the observed dispersion in photon indexes up to 100 keV \citep{nikolajuketal04a,wangetal13}.  The ``standard'' \citet{mathewsferland87} and \citep{koristaetal97}\ were also considered as suitable for Pop. A sources. The scatter in $\log(\kappa/\bar{\nu_{\rm i}})$\ derived from the values of the four continua is 0.033. We report this value in Table \ref{tab:budget}. 
%The actual scatter in the xA sample may  be lower  because xA sources are believed to show a very small dispersion in ionizing continuum properties . 
A more appropriate range is probably between the NLSy1 continuum   and the ``minimum'' spectral energy distribution (SED) of Fig. \ref{fig:contc}: if so, the scatter will be reduced to 0.02. This prediction should be tested by an analysis of   low-$z$\ sources for which X-ray data are available. 

\paragraph{Factor ${n_\mathrm{H}U}$ --} The term $ {n_\mathrm{H}U}$\ is only slightly affected by the frequency redistribution of ionizing continua (within the continua assumed).  However, $\log {n_\mathrm{H}U}$\ depends on the diagnostic ratios values. {\sc cloudy 13} simulations \citep{ferlandetal13} indicate  a difference of $\Delta \log {n_\mathrm{H}U} \approx 0.2$\ from ratio \aliii/\siiii $\approx$ 0.5 (A3$^\mathrm{m}$) and \aliii/\siiii $\approx$ 1.0 (A4$^\mathrm{m}$). Since we do not actually separate A3$^\mathrm{m}$\ and A4$^\mathrm{m}$\ sources, we consider the difference as a source of statistical error $\pm 0.1$\ around an average value. A proper treatment of the difference as a systematic effect will require a larger sample (\S \ref{system}). The measures of \citet{negreteetal13}\ on high S/N spectra indicate that measurement errors can be reduced to $\Delta \log {n_\mathrm{H}U} \approx 0.05$. 

%We derive $\sigma \approx$ 0.05 for the factor $\log \left[\kappa/(\bar{\nu_{\rm i}} 10^{n_\mathrm{H}U})\right]$. 

\paragraph{The structure factor $f_\mathrm{S}$} -- The value of $f_\mathrm{S}$\ is probably governed by   $\lambda_\mathrm{Edd}$\    \citep{netzermarziani10}.  An upper limit   variance was estimated by considering sources observed in reverberation mapping for which \mbh\ has been also derived from the \mbh\ -- bulge velocity dispersion relation \citep{onkenetal04}.  The resulting uncertainty $\delta f_\mathrm{S}/f_\mathrm{S} \approx 0.2$\ was  derived from an heterogeneous sample of sources whose emission line profiles and accretion rate are very different \citep{negreteetal13}. The \hb\  profiles of xA sources can be almost always modeled with an unshifted (virial) Lorentzian component plus an additional component affecting the line base. This  suggests the same, reproducible structure. We therefore expect a significantly smaller $\delta f_\mathrm{S}/f_\mathrm{S} $, $\rightarrow$ 0, if xA sources are considered as a distinct population in the analysis of the \mbh\ -- bulge velocity dispersion relation, and assume $\delta f_\mathrm{S}/f_\mathrm{S} = 0.1$\ in Col. 3 of Table \ref{tab:budget}. 

\paragraph{FWHM}--  Eq. \ref{eq:ln}  suggests that a most relevant source of statistical error may be $\delta v$\ that appears at the fourth power and  is assumed here to be represented by the FWHM of  \hb\  and intermediate  ionization lines \aliii\ and \siiii\ (\S \ref{sample}).   Indeed, a large fraction of the rms in the present sample can be accounted for by FWHM measurement errors,  especially in sample 3. In Table \ref{tab:budget} we conservatively estimate a typical error $\approx$15\%\ in FWHM measures. A lower, $\approx$ 5\%\ uncertainty in FWHM is within reach of dedicated observations.   

\subsubsection{$z$-based luminosity}

%In the following we will consider a mainly qualitative overview of sources of error that can affect the results. A quantitative analysis is deferred to further work. 
%Conversely ignoring the effect of orientation on line broadening  yields lower values of ``virial'' luminosity causing an overestimate of $H_{0}$.  

\paragraph{Aperture effects} -- Aperture losses and errors associated to the spectrophotometric calibration   should be carefully assessed. Here we conservatively set a 1-$\sigma$\ confidence error of 10\%\ following the SDSS website.\footnote{Errors in spectrophotometry 10\% of better: http://www.sdss.org/dr7/algorithms/spectrophotometry.html}

\paragraph{Redshift $z$} -- Detailed fitting can provide accurate redshift   making the  uncertainty on $z$\ negligible. However, $z$-values provided by the survey suffer from significant uncertainties \citep{hewettwild10}. We assume a $\delta z \approx 0.002$. This uncertainty   should be easily reduced to 0 by precise $z$\ measures unless quasars   show significant peculiar velocities with respect to the Hubble flow.  

\paragraph{Bolometric correction} -- The continuum shape has been represented here by a typical parameterization thought to be suitable for Pop. A sources \citep{mathewsferland87} that are moderate/high Eddington radiators, with a single bolometric correction  applied to all sources. The  dispersion around the assumed value has been estimated to be $\approx$ 20\%\ by \citet{richardsetal06} for quasars of all spectral types. The  dispersion around the typical spectral energy distribution of xA sources is probably lower.    The main argument for a small scatter is again that we are considering objects  thought to be producing similar emission line ratios. These ratios are dependent on the shape of the continuum \citep{negreteetal12}, making it  reasonable to assume that the continua of xA sources are similar and with a dispersion  significantly less than the one found in general surveys that do not distinguish spectral types. We derive a difference of $\approx$ 10\%\ between the NLSy1 SED and the minimum SED in Fig. \ref{fig:contc}.
%We conservatively adopt the value estimated by \citet{richardsetal06} in our error budget. The scatter between the SEDs considered a is $\approx$ 10\%, and this is a more likely value if restriction to xA source is made (Col. 3 of Table \ref{tab:budget}).

\paragraph{Continuum anisotropy} If the optical/UV continuum is emitted by an accretion disk, the observed luminosity should be dependent on the angle $\theta$\ between the disk axis and the line of sight, which will be different for each source. The angular dependence  can be written as $\lambda L_{\lambda} \approx \lambda L_{\lambda, \theta=0^\circ} \cos \theta (1+ a_{1} \cos\theta)$, where the second term is the limb-darkening effect \citep[e.g.,][in the case of a thin disk]{netzeretal92,netzertrakhtenbrot14}. 
The limb-darkening is wavelength dependent, and we are not aware of any exhaustive calculation  in the context of slim disks \citep{wataraietal01}. Nonetheless slim disks are optically thick thermal radiators so that a good starting point is to consider them  Lambertian radiators i.e., to { first} neglect the unknown but second order limb darkening effect, { and then apply a reasonable limb darkening term $a_{1} \approx 2$\   \citep{netzertrakhtenbrot14}}. We estimate that  anisotropy introduces a scatter of { $\delta \log L \approx 0.085 $}, if the probability of observing a source at $\theta$\ is $\propto \sin \theta$ with $0^{\circ} \le \theta \le 45^{\circ}$ in the case of a thin disk with no limb darkening effect. The scatter increases to $\delta \log L \la 0.15 $\ in the case of a slim disk  whose height is assumed to scale with radius for dimensionless accretion rate $\dot{m} = 1$  \citep{abramowiczetal88,abramowiczetal97}.

\subsubsection{Conclusion on statistical errors}
\label{budgetconcl}

The error sources described above and reported in  Col. 2  of Table \ref{tab:budget}\  provide a total error that accounts for and even  exceeds the observed rms. The observed dispersion $\approx$0.365\ implies that   dispersion values reported  in Table \ref{tab:budget} might be close to the lowest values.
Observational improvements can be devised  to reduce, whenever possible, the main source of statistical errors that remains the FWHM of virial broadening estimator.  Lowering FWHM measurements to $\approx$5\%\  would results in  rms $\approx $ 0.3. This rms values is obtainable without requiring advancements in our physical understanding and in the connection of Eddington ratio to observed properties. If the scatter in Eddington ratio could be reduced by accounting for systematic and random effects, then rms $\approx $  0.2 would become possible.

\subsection{Systematic errors}
\label{system}

\subsubsection{General considerations}

%We stress again that the determination of the $\Omega$s is affected  by statistical errors that could be significantly reduced with   better sample selection and improved data.  More explicitly,  criteria used for selecting  xA sources require that several parameters in Eq. \ref{eq:ln}  have the same values: $\kappa$ (\nh$U$), and $\bar{\nu_\mathrm{i}}$. Changing them  implies a change in line ratios.  Therefore,

Eq. \ref{eq:ln} is probably not influenced by systematic errors except for an offset  associated with the the assumed $\lambda_\mathrm{Edd}$, $\kappa$ (\nh$U$), and $\bar{\nu_\mathrm{i}}$. The offset will affect only $H_{0}$. There are also  systematic effects that could reduce $z$-derived source luminosities.  Light losses will also lead to systematic underestimates for luminosity which will affect  $H_{0}$.  Eq. \ref{eq:ln}  currently provides a valid redshift-independent luminosity estimator suited for the measure of \om\ and \ol\ only.

\subsubsection{Orientation bias}

{  Under the assumption  that observed line broadening $\delta v_\mathrm{obs}$\ is due to an isotropic component
$\delta v_\mathrm{iso}$ plus a planar Keplerian  component $ v_\mathrm{K}$,
the broadening can be expressed as $\delta v^{2}_\mathrm{obs} =
\delta v^{2}_\mathrm{iso} + v^{2}_\mathrm{K}\sin^{2}\theta$\
\citep[e.g.][]{collinetal06}. The probability of
observing a randomly-oriented source at $i$\ is $\sin \theta$, and the
systematic offset in Eddington ratio  can be  computed by integrating the
Eddington ratio values computed at $\delta v_\mathrm{obs}$ over their
probability of occurrence for a fixed $ v_\mathrm{K}$.
Assuming  $\delta v_\mathrm{iso}/\delta v_\mathrm{K} = \frac{1}{2}$, $5
\ltsim \theta \ltsim 45$\ the offset in \lledd\ is approximately a factor of 2.
Therefore, the Eddington ratio we are considering (and whose distribution
is shown in Fig. \ref{fig:lleddour}) could be subject to a substantial
bias. However, since  orientation effects are not included in
$\lambda_\mathrm{Edd}$\ computations and $\lambda_\mathrm{Edd} \propto
1/v^{2}$,  a correction to the virial broadening estimator in Eq.
\ref{eq:ln} would be compensated by the change in $\lambda_\mathrm{Edd}$,
leading to a net   $\Delta \log L \approx 0.0$. }

In this paper  the  $\lambda_\mathrm{Edd}$\ dispersion reported in Tab.
\ref{tab:budget} already takes into account orientation effects since
$\delta \lambda_\mathrm{Edd}$\ has been estimated  from the a  posteriori
computed distribution of $\lambda_\mathrm{Edd}$ \ that is broadened by the
orientation effect.  A more proper approach would be to
derive an orientation angle for individual sources. xA sources likely
possess strong radiatively-driven winds whose physics and effect on the
line profile of a high-ionization line like \civ\ can be modeled
\citep{murraychiang97,progaetal00,risalitielvis10,flohicetal12}.

\subsubsection{Major expected systematic effects  { as a function of $z$, $L$, and sample selection criteria}}

 {In the 4DE1 approach luminosity dependences are in the parameters correlated with  Eigenvector 2 \citep{borosongreen92}. The main correlate is the equivalent width of \civ\ (i.e., the well known ``Baldwin effect''; \citealt[e.g.,][and references therein]{bianetal12}). In that case, indeed, even if the most likely explanation of the decrease of W(\civ) with luminosity found in large sample is a dependence of W(\civ) on Eddington ratio plus selection effects in flux limited sample \citep{baskinlaor04,bachevetal04,marzianietal08a}, the \civ\ equivalent width should be avoided  because of a possible residual dependence with luminosity. The diagnostic ratios we employ are meant to minimize any possible dependence from a physical parameter that is in turn related to luminosity, like for example the covering fraction of the emitting gas that may affect W(\civ). } 

Neither \rfe\ nor \aliii/\siiii\ depend significantly on $z$ or luminosity in xA sources. Selecting xA
sources from the \citet{shenetal11} sample, correlation coefficients are
not significantly different from 0. For 378 sources with \rfe$\ge$1, the
Pearson correlation coefficients are 0.075 and 0.09 with $z$ and $L$\
respectively. These values are both not significant even if the   sample of
sources is relatively large.  The  \aliii/\siiii\ ratio  as a function of
redshift and luminosity   yields    { least-square}  fittings lines that not significantly different from 0. { We have verified that the optical and UV selection criteria are mutually consistent in \S \ \ref{consist}, and shown that the conditions on \rfe\ and \aliii\ -- \ciii\ are simultaneously satisfied over a wide range $L$ and $z$, even if  for  the only 9 xA  sources with available data: from 1H 0707-495 (the lowest luminosity A3 source, $\log L\approx  44.6$\ [\ergss] at $z \approx 0.04$) to  HE1347-2457 ($\log L\approx  47.9$, at $z \approx  2.599$).  } 

{ Residual systematic effects can arise if the quantities entering Eq. \ref{eq:ln} are  different in high- and low-$z$ sources i.e., in \rfe-selected and 1900 -selected samples.} 

\begin{description}
\item[ \lledd --- ] Our present $N = $92 sample   suffers from a slight \lledd\ bias.   The difference in \lledd\ between sample 1 and 3 (optical and UV based)  indicates that $\delta \log$ \lledd $\la 0.08$.  

{  Several   works suggest a relation  between diagnostic ratios \rfe,  \aliii/ \siiii\ and \siiii/\ciii, and   Eddington ratio    \citep{marzianietal03b,shenetal11,aokiyoshida99,willsetal99}.   There is no  correlation at all between \rfe\ and the UV ratios  employed in this paper as far as the general population of AGNs is concerned (Fig. \ref{fig:test}). However this may not be longer true when \rfe $\rightarrow 1$: if \rfe$\approx$1, \aliii/\siiii$\approx$0.5\ and \ciii/\siiii$\approx$1, while for larger \rfe\ \aliii\ becomes stronger and \ciii\ weaker. Indeed,  4DE1  suggests a relation between \lledd\ and metallicity,   with the most metal rich system being associated only with the sources accreting at the highest rate.  An important  result of the present  investigation is right that xA sources -- that are  extremely metal rich \citep{negreteetal12} -- are revealed over a broad range of $L$ and $z$, but only for a very narrow range of \lledd. As stressed earlier, luminosity and $z$\ cannot be major correlates. %A correlation between  luminosity and metallicity \citep{hamannferland93,juarezetal09,shinetal13} is expected to be mainly the result of a Malmquist bias, since at high $z$ and high $L$\ sources radiating at higher \lledd\ are preferentially selected. 

As long as the distribution of  the measured intensity ratios \aliii/\siiii, \siiii/\ciii\ and \rfe\ are independent of $z$\  and $L$\ (absence of even a weak correlation could be enforced by resampling to avoid small effects),   a relation (if any) between \rfe\ and \aliii/\siiii, \siiii/\ciii\ is established,  inter-sample systematic effects should be minimized. The 9 sources considered in \S \ref{consist}\ are clearly insufficient to test these conditions.  A proper observational strategy  could involve: (1) a large  sample  to define  a relation between \rfe\ and \aliii/\siii, \siiii/\ciii\  (and \lledd). This is especially needed since any $\lambda_\mathrm{Edd}$  deviation as a function of redshift can introduce a significant systematic effect: for example,    $\delta \log$\lledd $\approx -0.05$\ between sample 3 and 1  implies $\delta $\om $\approx$ 0.05; (2) a vetted subsample with \hb\ observations in  the near IR and simultaneous 1900 observations in the optical.  }

%It is therefore advisable to   test/establish any relation between \rfe\ and the UV ratios for highly accreting sources. 

\item[$\kappa/ h\bar{\nu}$ ---] { A change of ionizing
continuum shape as a function of redshift and/or luminosity,  namely of
the ratio $\kappa/ h\bar{\nu}_i$\  for a fixed \aliii/ \siiii, \siiii/ \ciii\ and \rfe. 
This could occur, practically, if we were selecting more and more extreme objects with the steepest continua.} The properties of the ionizing continuum in quasars are not very well known; however,
it is reasonable to assume that Pop. A sources show continua intermediate
between the continuua shown in Fig. \ref{fig:contc}. If the  continua
labeled as \citet{koristaetal97} and the extreme, minimum NLSy1 continuum
(dashed line)  are considered, $\kappa/h\bar{\nu}_i$ will change from
0.174 to 0.207, with a   $\delta \log L \approx 0.075$.    xA
sources are expected to show a steep X-ray continuum
\citep{grupeetal10,panessaetal11,wangetal13}, so that the range of change
may be more realistically bracketed by the typical NLSy1 and minimum
continuum. In this case, the hypothesis of a systematic evolution implies a
change in $\kappa/h\bar{\nu}_i$ that leads to $\delta \log L \approx
0.035$.    { However, the change in \nh$U$\ associated with  different
continua  {(for the same values of the observed \aliii/\siii\ ratios)} tends to compensate for the change in  $\kappa/h\bar{\nu}_i$, reducing the effect to $\delta \log L \approx 0.026$}.  There is no
evidence for such evolution/selection effects in xA sources, but the relevance of systematic
continuum changes and the relation between  $\kappa$ (\nh$U$), and $\bar{\nu_\mathrm{i}}$\ and diagnostic ratios 
should be tested with dedicated X-ray observations of
 high-$z$\ xA sources. 

\item[Ionizing photon flux \nh$U$ --- ] { The factor  \nh$U$ is set by the   \aliii/ \siiii, \siiii/ \ciii\ and \rfe\ ratios and, for a given continuum shape, we do not expect systematic changes. When selecting large  samples some degree of heterogeneity is unavoidable.  Differences in the Eq. \ref{eq:ln} parameter  values for bins A3$^\mathrm{m}$\ and A4$^\mathrm{m}$ will contribute  to the overall sample rms but should not introduce any systematic effects as long as the fraction A3$^\mathrm{m}$/A4$^\mathrm{m}$\ or, more properly,  the  \aliii/ \siiii, \siiii/ \ciii\ and \rfe\   distributions   are consistent and independent from $z$.   }

\item[Structure factor $f_\mathrm{S}$] --  { The self-similarity of the profiles over $z$ and luminosity should be carefully tested on high S/N spectra since the FWHM of the lines is clearly dependent on the assumed profile shape. Available data  support the assumption that the shape is not changing as a function of FWHM and $z$\ (as  also shown by  the data of  Fig.  \ref{fig:examphb} and Fig.  \ref{fig:exampaliii}): a Lorentzian function yielded good fits for   \hb, \aliii, \siii\ in all cases considered in this paper. This is also true when fits are made to carefully selected composite spectra with S/N $\ga$100 \citep[see e.g.,][]{zamfiretal10}.}

%between sample 1 and 3    \ref{fig:lleddour} :

\end{description}

An important systematic effect is related to the use of equation determination of the quasar luminosity from redshift (Eq. \ref{eq:zlum}).

\begin{description}
  \item[$B. C.(1800) -- B. C.(5100) $ ---] The ratio between the bolometric correction at 5100 \AA\ and 1800 \AA\ has been assumed in this paper to be $\lambda f_\lambda(5100) / \lambda f_\lambda(1800)$\ = 0.63 on the basis of the \citet{mathewsferland87} continuum. A more proper value could be well 0.7, as measured on the optical and UV spectral of I Zw1. This change will shift the  estimate of \om\ by $\approx$ 0.05, to \om$\approx$ 0.25, in the case of constrained \om + \ol = 1.0. In addition if $\lambda f_\lambda(5100) / \lambda f_\lambda(1800) \approx $\   0.7 the systematic offset in the \lledd\ distribution between sample 1 and 3 will be reduced by $\approx$ 0.05.  The average ratio  $\lambda f_\lambda(5100) / \lambda f_\lambda(1800)$\   and the associated dispersion (and hence $B.C.(1800) -- B. C.(5100) $)\ should be carefully  established by dedicated observations. Such observations are also needed because continuum anisotropy leads, in addition to a random error,   to a systematic underestimate of   source luminosity, as discussed below. The effect is relevant here as long as it is wavelength dependent if optical and UV data are considered together. Recent observational work on radio-loud sources suggests that the degree of anisotropy changes very little with wavelength  \citep{runnoeetal13}.  However, this may not be the case for slim disks of highly accreting sources. 
\end{description}

The  most relevant  systematic effects, among the ones that are listed above,  are related to  $B. C.(1800)/B. C.(5100) $ and to the distributions of the optical and UV line ratios that could be linked to small -- but significant -- trends in \lledd. We have  shown that other effects, like evolution of the ionising continuum -- within reasonable limits --  may yield  a modest systematic effect on luminosity estimates, $\approx$ 0.03 dex. 

The following effects  are more speculative in nature and of lower relevance.

\begin{itemize}
 %This assumes that the factor related to the ionizing photon flux \nh$U$ remains constant. 

%\item{Line broadening. $L$ is dependent to the  FWHM (or any virial broadening estimator, like line velocity dispersion) with the fourth power of luminosity. Great care should therefore be exerted not only on minimising statistical errors, but also on the choice of the fitting function. Our preliminary analysis constantly assumed a Lorentzian function for all lines, a choice appropriate for Pop. A sources indicated by several previous studies \citep[e.g.,][and references therein]{negreteetal13}. A change in the fitting function  will change significantly the FWHM estimate (for example, by $\approx$ 25\%\ using a Gaussian). There is no evidence that the shape of the best fitting function changes either with redshift or luminosity, although higher S/N spectra may reveal a more complex scenario.   }

\item { Continuum anisotropy is also expected to give rise to a
Malmquist-type bias in $z$\ based luminosities. Close to a survey limiting
magnitude the brightest (face-on) sources will be selected preferentially.
Unlike relativistic beaming however,   disk anisotropy is of relatively
modest amplitude. In the case of a slim disk with limb darkening the
difference between a face-on source and a randomly oriented sample of
objects is $\delta \log L \approx 0.2$ dex. Since core-dominated
radio-loud sources are expected to have an additional relativistically
beamed synchrotron continuum component, they should be avoided from any
sample. If sources are selected from a large flux limited sample, (1) either orientation is inferred from \civ\ line profile modelling, or (2) a  more elementary precaution would be to consider sources brighter than $\approx 0.5$\ { mag} than their discovery survey (i.e., the SDSS in the case of sample 3 limiting magnitude).}

\item{ Intervening large scale structures are expected to produce a
lensing effect on the light emitted by distant quasars. This effect is
noticeable especially for sources at $z \ga$1
\citep{holzwald98,holzlinder05}. The lensing effect is however found to 
be averaged out for large samples ($\la 100$\ sources) which is the case
of any quasar sample that could be realistically employed
for cosmology. }
\end{itemize}

The present samples hint  at (small) systematic differences that can be more clearly revealed and quantified only with a larger sample and/or vetted.   {   As long as we employ the same diagnostic ratios (with consistent distributions of values as a function of $z$) and the same line profile model, residual systematic effects with $z$, $L$, and \lledd  should be minimized.  Any systematic change with $z$ and $L$\  affecting  the parameters entering into Eq. \ref{eq:ln}  will also affect the intensity ratios and will become detectable. }{ Assessing and} avoiding systematic effects would  require  a uniform redshift coverage, as assumed for the   mock samples. More details on a possible observational strategy are give in \S \ref{strats}.

%A further tool to identify A3 and A4 sources is the peculiar \mgii\ profile in these sources \citep{marzianietal13}, characterized by blueshifts of a few hundred \kms. The narrow  \oii\ doublet and the \mgii\ line are easily covered in the same spectrum if $0.4 \la $z$\ \la 1.4$. This technique could be applied especially between $0.9 \la $z$\ \la 1.4$\ where \hb\ coverage is more difficult.  %\paragraph{Product (\nh $U$)} Shape of the continuum and product ($n U$) are related. Even, as noted before, changes in continuum and physical parameters should be self compensating, the influence of ionizing spectral energy distribution  on physical parameters should be further investigated. 
%\paragraph{Eddington ratio $\lambda_\mathrm{Edd}$} 
%\paragraph{Geometry factor} The factor $f$\ is assumed as derived for Pop. A ($\approx$ 2) by \citet{collinetal06} but its determination on an object by object basis is an open and  challenging problem \citep{netzermarziani10}. Our assumptions on emission line ratios should yield to a well-defined value but further work is needed to confirm that $f$ is actually $\approx$ 2 for xA sources. 
%On the other hand, an observational strategy based on the selection of a gold sample of the most extreme sources (i.e., A4) with fewer sources and presumably smaller scatter may also be possible.

\section{Discussion}
\label{disc}

The idea to use quasars as Eddington standard candles is not new
\citep[e.g.,][]{marzianietal03d,teerikorpi05,bartelmannetal09,sulenticetal12a,wangetal13}.
Luminosity correlations were the past great hope for using quasars as standard candles.
The most promising  luminosity correlation involved the Baldwin effect  
 \citep{baldwin77,baldwinetal78} which is now thought to be governed by
Eddington ratio \citep{bachevetal04,baskinlaor04}. In any case it is too
weak to provide interesting cosmological constraints.  Other methods using line width measures have also been proposed \citep[e.g.,][]{rudgeraine99} although it is unclear 
that the line width distribution shows real change with redshift. Recently, a somewhat similar proposal to the one presented in this paper has been advanced which advocates the hard X-ray spectra index 
as a selector for extreme sources \citep{wangetal13}. We considered both hard and soft X-ray 
measures when selecting the principal 4DE1 parameters and concluded that hard measures 
showed too little dispersion across the 4DE1 optical plane  compared to $\Gamma_\mathrm{soft}$ \citep{sulenticetal00b}. 

%Eq. \ref{eq:ln} appears robust as far as estimation of \om\ and \ol\ are concerned since, in this case, the necessary condition is that there is no parameter dependence/evolution with $L$\ and $z$ entering Eq. \ref{eq:ln}. This condition is consistent with observations of sources whose diagnostic line ratios in the optical and UV are very similar over the large range of $z$\ and $L$\ considered here. The use of Eq. \ref{eq:ln}  is more problematic for  determination of $H_{0}$\ since several parameters are poorly known (especially $f_\mathrm{S}$\ and $\lambda_\mathrm{Edd}$). 

The  concordance $H_{0}$\ value has to be assigned a priori to constrain \om\ and \ol\   to avoid circularity, { unless the normalised slope is used as a best-fit estimator. } The determination of $H_{0}$\ is not fully independent of $z$-based distances. The Eddington ratio estimate   for I Zw 1  requires a luminosity computation that assumes a value of $H_{0}$.  
%Most quasars included in this study are at least  $\sim$10 times  more distant than I Zw 1. We assume that all sources showing a I Zw 1-type spectrum radiate at extreme  Eddington ratios (assumed to be unity).  The actual value of $\lambda_\mathrm{Edd}$\ could be different if systematic  effects are considered (see below). 
In order to make an independent determination of $\lambda_\mathrm{Edd}$\ (and 
hence of \h) at least one redshift independent luminosity determination would be needed (e.g. 
distance inferred from a type Ia supernova). In this case \mbh\ would follow from luminosity via
the virial relation with $r \propto L^{1/2}$ and \lledd\ from the ratio $L$/\mbh. Another approach would 
be to derive $\lambda_\mathrm{Edd}$\ from a physical model of the high-ionization outflow common to extreme Eddington  sources. 

The definition of ionization parameter also involves $L$. However
ionization parameter and density values were derived from
emission line ratios. The luminosity is a theoretical luminosity that, for
an assumed continuum, yields the number of ionizing photons needed to
produce the emission lines. We retrieve $Un_\mathrm{H}$\ from emission line ratios;
no flux or line luminosity measurements are involved. The assumption that
$r \propto L^{0.5}$ is consistent with the assumption of $Un_\mathrm{H}$ = const
(actually, it follows from the definition of $U$). Therefore, Eq.
\ref{eq:ln} has no implicit circularity. 

%An important assumption in this approach is that the virial relation is 
%valid even at high \lledd $\rightarrow$ 1. It could well be valid for very dense 
%gas associated with the Balmer and intermediate ionization emission lines
%\citep{netzermarziani10,marzianietal10,negreteetal12}. Profiles of Balmer
%lines are usually unshifted and symmetric suggesting that radial/partly
%obscured flows are not strongly affecting them -- at least not at the FWHM level
%where line width is usually measured. As mentioned earlier,
%FWHM values are characterised by a well-defined curve \citep{marzianietal09} suggesting 
%a systematic increase of minimum observed FWHM with luminosity. This relation can
%be straight forwardly explained if $r \propto L^{0.5}$, and the virial
%relation holds at \lledd = 1. Actually it is remarkable for our
%understanding of quasars that FWHM of an emission lines like \mgii\ is
%consistent with the virial broadening derived from \hb. This consistency
%is apparently maintained for objects at $z\approx$3. This result facilitates
%single epoch \mbh\ computation at high $z$\ quasars.

\subsection{Comparison with the Supernova Legacy Survey}

There are many analogies between supernova surveys \citep{guyetal10} and our proposed 
method using quasars. Both methods rely on intrinsic luminosity estimates for a large number of 
discrete sources. The advantage of the supernova surveys is that individual supernov\ae\ show 
a smaller  scatter in luminosity \citep[e.g.,][]{riessetal01}. However, very few supernov\ae\ have 
been detected at $z \ga 1$\ while a quasars sample can be easily extended (with significant 
numbers) to $z  \approx 3$\ or possibly  $z  \approx  4$. Differences in redshift coverage account 
for the different sensitivity to  $\Lambda$: \ol\ is tightly constrained using supernov\ae\ while it 
remains loosely constrained using quasars. Quasars are distributed over a redshift range where \om\ 
ruled the expansion of the Universe while supernov\ae\ sample epochs of accelerated expansion 
(Fig. \ref{fig:diff}).  But quasars can sample any range covered by supernov\ae. Statistical errors for \om\ derived from the mock samples (with unconstrained \om\ and \ol) are lower than ones derived from the first three 
years of the supernova legacy survey that yield \om $\approx 0.19^{+0.08}_{-0.10}$\ \citep{conleyetal11}.

\subsection{Possible observational strategies}
\label{strats}

In order to exploit a sample of high \lledd\ radiators both calibration
observations and a larger sample of extreme Eddington sources are needed.
Simultaneous rest frame UV and optical observations covering the 1900 and \hb\ 
range are needed (a feat within the reach of present day multi-branch spectrometers): (1) to constrain the bolometric corrections and specifically the 
$B. C.(1800)/B. C.(5100) $  ratio; (2) to define systematic differences between spectral 
types A3$^\mathrm{m}$\ and A4$^\mathrm{m}$, including the $B.C.$ An attempt should  be also made to cover with a large  \hb\ sample the $z$ range 0.1  -- 1.5, where the effect of a nonzero \ol\ is 
most noticeable and where any rest-frame optical/UV inter calibration is not   needed. { A related option is   to obtain \rfe\  only, covering the redshifted \hb\ spectral range into the near and mid IR ($K$ band observations can reach $z \approx 3.5$).} { Alternatively, the UV 1900 blend can be easily covered by optical spectrometers over the redshift range 1.1 $\la z \le 3.5$. This approach would allow to measure \om\ without the encumbrance of an inter calibration with \hb\  data. }

%A first step could involve to define $L \approx \mathcal{L}_{0} \delta v^{4}$\ with $\mathcal{L}_{0}$\ dependent on spectral type.
%; (3) to further verify that FWHM \hb\ and \aliii\ are equally affected by virial
%broadening. 
% Archival data in the X-ray domain can be collected for a subsample in order
 %to estimate an appropriate value of and its dispersion. 
 
The best hope for accurate and precise results rests in a ``brute
force'' application of the method to a large sample. Tab. \ref{tab:budget}
shows that rms$\approx$ 0.3 can be obtained with better data. With
uniform redshift coverage the precision of the method will scale with
rms/$\sqrt{N}$. This means that a precision similar to the one obtained
with the mock sample (rms = 0.2) can be achieved with a sample of 400
sources  while a precision similar to rms = 0.1 would require a sample
near 1000 quasars. It is possible  that a sample of this size (or even
larger) can be selected from spectra collected by recent major optical 
surveys \citep[e.g., ][]{parisetal12}.  
%A strategy based on selection of a smaller sample of the most extreme 
%sources (i.e., bin A4) likely with smaller scatter may also be possible. 
%Selection and sample refinement employing all of the 4DE1 parameters 
%would require many more (optical, IR and X-ray) observations that 
%would likely yield a smaller but more constraining sample. 
%We estimate that observations for calibration and S/N improvement towards  
%a 400 source sample would require $\sim 100$\ hours observing time on 
%large aperture optical telescopes. 

\section{Conclusion}

%The problem of $z$-independent luminosity estimation is shifted to the virial \mbh\
%estimation which is relatively standard and accurate. 
We have shown that sources radiating at,
or close to,  \lledd $\approx$1 can be identified  in significant numbers with reasonable confidence.
 { These sources show stable emission line ratios over a very wide range of $z$ and $L$. We have performed exploratory computations and shown that these sources -- apart from their intrinsic importance for quasar physics -- may be also  prime candidates as cosmological probes.} We then presented a method for using some quasars as redshift-independent distance estimators.  We do not claim to present constraining results in this paper beyond showing an overall 
consistency with concordance cosmology and exclusion of extreme  models (e.g. a flat Universe 
dominated by the cosmological constant)  starting from an estimates of the most likely values of quasar physical parameters entering in  Eq. \ref{eq:ln}.  Our goal was to identify suitable quasars, to describe a possible  approach capable of yielding meaningful constraints on \om\ and \ol,  and to identify most serious statistical and systematic sources of uncertainty.   A quantitative analysis of systematic effects due to continuum shape, orientation,    $f_\mathrm{S}$\  and $\lambda_\mathrm{Edd}$    as well as an attempt at reducing statistical errors is deferred to further work. Addressing and overcoming   systematic biases  requires dedicated, but feasible, 
new observations. 

\begin{figure}
%\epsscale{.75}
%\includegraphics[scale=0.3]{mocksample100_0.1.eps}\vspace{3cm}\includegraphics[scale=0.3]{h0ommedia100_0.1.eps}\\
%\includegraphics[scale=0.3]{slope100_0.1.eps}\\
\hspace{-0.cm}\includegraphics[scale=0.5,angle=-90]{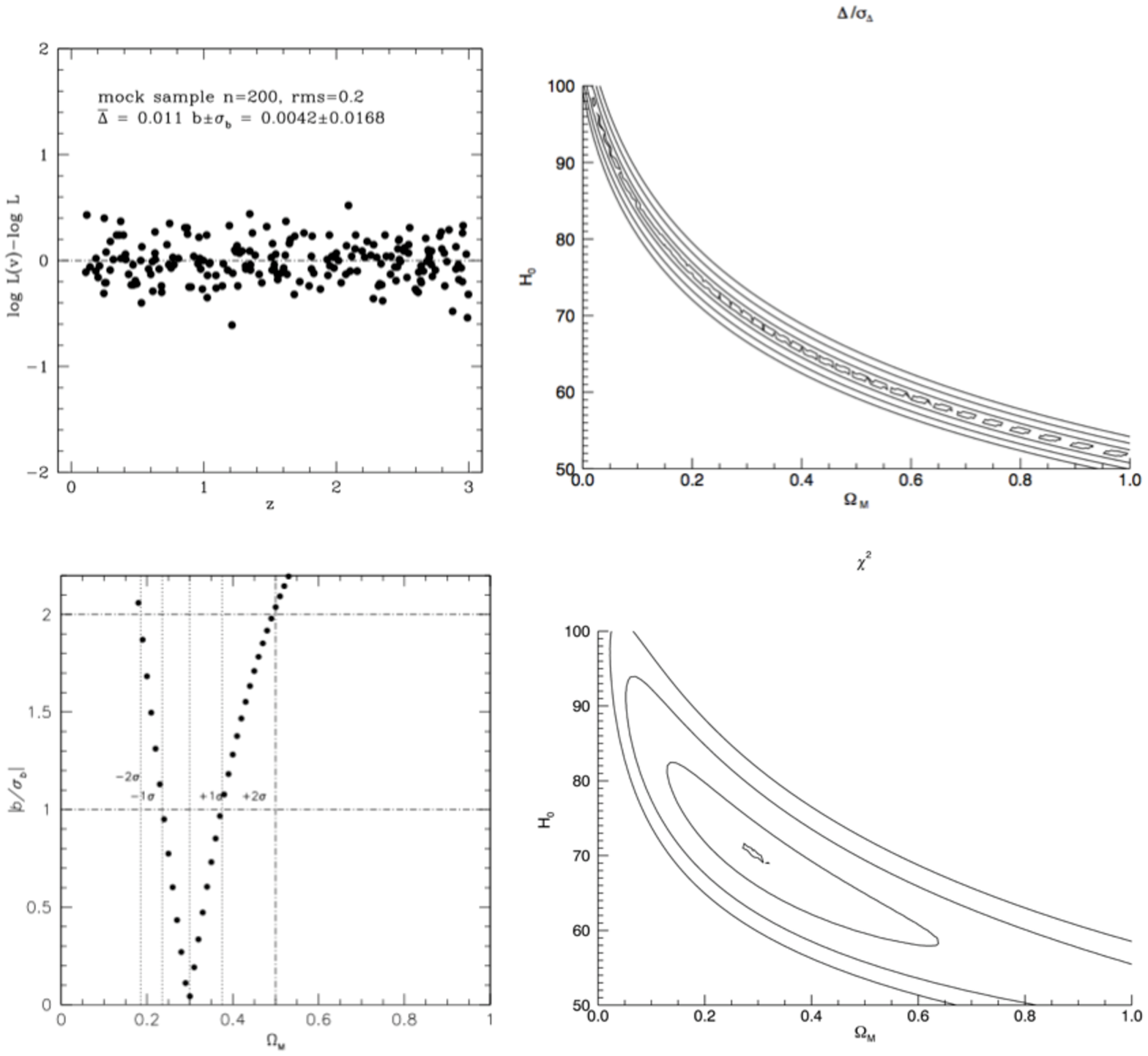}
\caption{Hypothetical results on mock sample with  rms=0.2.  The  upper left panel shows 200 synthetic data points whose virial luminosity is  assumed to deviate (randomly) from the concordance case according to a Gaussian distribution with $\sigma$ = 0.2.  Upper right:  $\bar{\Delta}/\sigma_{\bar{\Delta}}$ as a function of \h\ and \om.  { Bottom left:  $|b/\sigma_\mathrm{b}|$\ as a function of \om. }   Lower right: normalized $\chi^2$\ as a function of \h\ and \om. $\bar{\Delta}/\sigma_{\bar{\Delta}}$, $|b/\sigma_\mathrm{b}|$\ and $\chi^2$\ computations  assume a  a flat geometry ($\Omega_\mathrm{M} + \Omega_{\Lambda} =1.00$). See text for more details. 
\label{fig:mock200}}
\end{figure}

\begin{figure}
%\epsscale{.75}
%\includegraphics[scale=0.3]{mocksample200_0.2.eps}
%\includegraphics[scale=0.53,trim=100 2000 100  2000 ,clip]{h0ommedia200_0.2.eps}\\
%\includegraphics[scale=0.3]{slope200_0.2.eps}\\
%\includegraphics[scale=0.4]{h0om100_0.1.eps}
%\includegraphics[scale=0.3]{chifree200_0.2.eps}\\
\hspace{0cm}\includegraphics[scale=0.5,angle=-90]{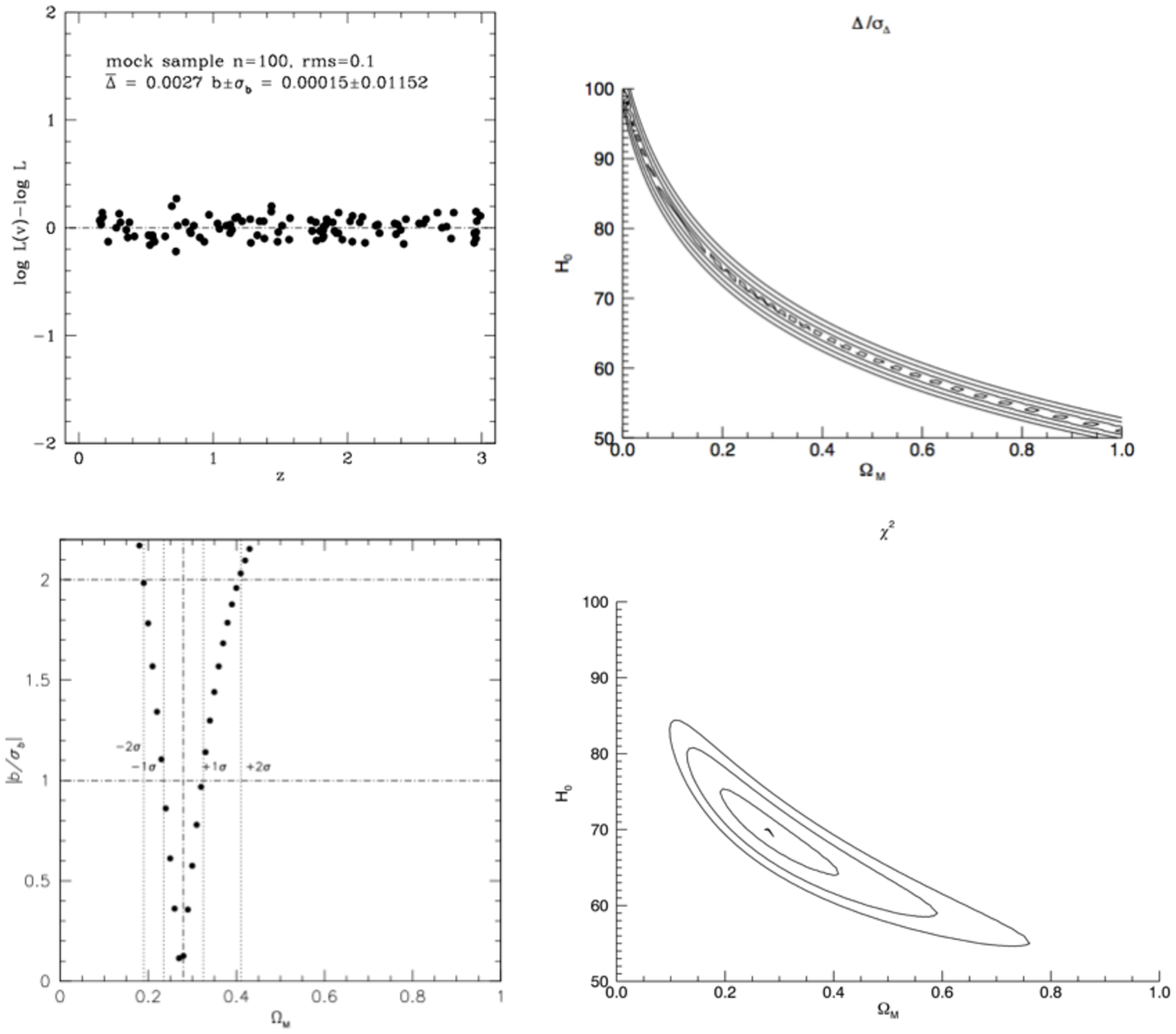}
\caption{Hypothetical results on mock sample with rms=0.1. Meaning of panels and symbols is the same of  previous Figure. %In the bottom right panel the lines trace $\chi^{2}/\chi^{2}_\mathrm{min} \approx $1.01, 1.07, 1.27. The latter two values correspond  to confidence limits of 1 and 2 $\sigma$\ respectively.   
\label{fig:mock100}}
\end{figure}

We stress that the precision of 
our method can be greatly improved with high S/N spectroscopic observations for significant 
samples of quasars. This is not just the usual refrain claiming that improvement in S/N can lead to
unspecified advancements: previous work shows that broad-line FWHM for Pop. A sources 
can be measured with a typical accuracy of 10\%\ at  a 2$\sigma$\ confidence level.
This would represent a major improvement with respect to the 20\%\ at 1$\sigma$\ uncertainty for 
many FWHM values used in this work -- implying 
$\Delta \log L $\ errors  decreasing from $\approx 0.7$ to $\ltsim 0.1$. Figs.
\ref{fig:mock200} and \ref{fig:mock100} shows that cosmologically meaningful 
limits can be set even with currently obtainable data. 

% The method presented in this paper could bring meaningful results relatively
%quickly without additional technological advancements.

%\acknowledgments
\bigskip
PM and JS acknowledge 
support from la Junta de Andaluc\'{\i}a, through grant TIC-114, Proyectos de Excelencias 
P08-FQM-04205  + P08-TIC-3531 as well as the Spanish Ministry for Science and Innovation through grant AYA2010-15169.  

Funding for the SDSS and SDSS-II has been provided by the Alfred P. Sloan Foundation, the Participating Institutions, the National Science Foundation, the U.S. Department of Energy, the National Aeronautics and Space Administration, the Japanese Monbukagakusho, the Max Planck Society, and the Higher Education Funding Council for England. The SDSS Web Site is http://www.sdss.org/.

The SDSS is managed by the Astrophysical Research Consortium for the Participating Institutions. The Participating Institutions are the American Museum of Natural History, Astrophysical Institute Potsdam, University of Basel, University of Cambridge, Case Western Reserve University, University of Chicago, Drexel University, Fermilab, the Institute for Advanced Study, the Japan Participation Group, Johns Hopkins University, the Joint Institute for Nuclear Astrophysics, the Kavli Institute for Particle Astrophysics and Cosmology, the Korean Scientist Group, the Chinese Academy of Sciences (LAMOST), Los Alamos National Laboratory, the Max-Planck-Institute for Astronomy (MPIA), the Max-Planck-Institute for Astrophysics (MPA), New Mexico State University, Ohio State University, University of Pittsburgh, University of Portsmouth, Princeton University, the United States Naval Observatory, and the University of Washington.

\clearpage

\bibliographystyle{apj} 
%\bibliography{biblioletter}

\clearpage

\clearpage

\hoffset=-1cm
\begin{table}
\begin{center}
\caption{Measured quantities for the \hb\ based samples \label{tab:hb}}
\setlength{\tabcolsep}{5pt}
\begin{tabular}{lcccccc}\hline\hline
\multicolumn{1}{c}{Source} &   $z^{\mathrm{a}}$     &  $\lambda f_{\lambda}^\mathrm{b}$ & S/N & FWHM& $\delta $FWHM$^{\mathrm{c}}$ & Sample   \\ 
& &  [\ergss \cmq] & & [\kms] & [\kms]&\\
\hline 
SDSSJ014247.74-084517.5	&	0.571	&	3.23E-12	&	15	&	2230	&	220	&	1	\\	
SDSSJ030000.01-080356.9	&	0.562	&	7.31E-12	&	36	&	2100	&	210	&	1a	 	\\
SDSSJ074840.52+154456.8	&	0.469	&	4.50E-12	&	25	&	2450	&	250	&	1	\\	
SDSSJ080908.14+461925.6	&	0.657	&	6.34E-12	&	23	&	2290	&	230	&	1	\\	
SDSSJ082024.22+233450.4	&	0.470	&	3.12E-12	&	25	&	1720	&	170	&	1	\\	
SDSSJ085557.11+561534.7	&	0.715	&	6.49E-12	&	25	&	2680	&	270	&	1a	 	\\
SDSSJ090423.31+400704.7	&	0.410	&	4.27E-12	&	20	&	1870	&	190	&	1	\\	
SDSSJ090840.71+132117.3	&	0.458	&	2.40E-12	&	16	&	2370	&	240	&	1	\\	
SDSSJ093531.61+354101.0	&	0.494	&	4.58E-12	&	18	&	2140	&	210	&	1a	 	\\
SDSSJ094033.76+462315.0	&	0.696	&	4.58E-12	&	15	&	2010	&	200	&	1	\\	
SDSSJ103457.27+235638.1	&	0.419	&	2.39E-12	&	21	&	1420	&	140	&	1	\\	
SDSSJ104613.73+525554.2	&	0.503	&	2.94E-12	&	22	&	1790	&	180	&	1	\\	
SDSSJ104817.98+312905.8	&	0.452	&	5.19E-12	&	20	&	1870	&	190	&	1	\\	
SDSSJ105205.58+364039.6 	&	0.609	&	5.76E-12	&	35	&	2330	&	230	&	1a	 	\\
SDSSJ105600.08+142411.3	&	0.623	&	3.94E-12	&	24	&	2500	&	250	&	1	\\	
SDSSJ110312.93+414154.9	&	0.402	&	1.01E-11	&	34	&	2050	&	210	&	1	\\	
SDSSJ110406.94+314111.5	&	0.434	&	5.02E-12	&	36	&	2030	&	200	&	1a	 	\\
SDSSJ111909.51+153216.4	&	0.674	&	5.23E-12	&	24	&	2800	&	280	&	1a	 	\\
SDSSJ112756.76+115427.1	&	0.510	&	4.76E-12	&	16	&	1820	&	180	&	1	\\	
SDSSJ112757.41+644118.4	&	0.695	&	4.03E-12	&	17	&	1850	&	190	&	1	\\	
SDSSJ113338.64+220026.8	&	0.546	&	2.78E-12	&	17	&	2080	&	210	&	1	\\	
SDSSJ113625.42+100523.2	&	0.552	&	3.13E-12	&	20	&	1460	&	150	&	1	\\	
SDSSJ113944.64+121436.0	&	0.618	&	3.01E-12	&	17	&	1420	&	140	&	1	\\	
SDSSJ120633.07+412536.1	&	0.554	&	4.87E-12	&	21	&	2110	&	210	&	1a	 	\\
SDSSJ120734.63+150643.6	&	0.750	&	4.51E-12	&	19	&	2240	&	220	&	1	\\	
SDSSJ121850.52+101554.1	&	0.543	&	4.06E-12	&	25	&	1700	&	170	&	1a	 	\\
SDSSJ122557.86+364907.7	&	0.477	&	3.77E-12	&	22	&	2210	&	220	&	1	\\	
SDSSJ124511.25+335610.1	&	0.711	&	7.21E-12	&	20	&	2270	&	230	&	1	\\	
SDSSJ130112.93+590206.7	&	0.476	&	1.66E-11	&	42	&	2250	&	230	&	1	\\	
	\hline\end{tabular}
\begin{list}{}{}
\item[$^\mathrm{a}$]{Redshift provided in the SDSS file header.}
\item[$^\mathrm{b}$]{At 5100 \AA\ rest frame.}
\item[$^\mathrm{c}$]{1 $\sigma$\ confidence level.}
%\item[$^\mathrm{d}$]{S/N at 5100 }
\end{list}
\end{center}
\end{table}

\clearpage

\addtocounter{table}{-1}

\hoffset=-1cm
\begin{table*}
\begin{center}
\caption{(cont.) Measured quantities for the \hb\ based samples}
\setlength{\tabcolsep}{5pt}
\begin{tabular}{lcccccc}\hline\hline
\multicolumn{1}{c}{Source} &   $z^{\mathrm{a}}$     &  $\lambda f_{\lambda}^\mathrm{b}$ & S/N & FWHM& $\delta $FWHM$^{\mathrm{c}}$ & Sample   \\ 
& &  [\ergss \cmq] & & [\kms] & [\kms]&\\
\hline 
SDSSJ130357.42+103313.5	&	0.589	&	3.87E-12	&	21	&	2210	&	220	&	1	\\	
SDSSJ132048.67+510313.7	&	0.466	&	4.74E-12	&	48	&	2040	&	200	&	1a	 	\\
SDSSJ133225.99+151926.4	&	0.466	&	3.55E-12	&	24	&	2160	&	220	&	1	\\	
SDSSJ133602.01+172513.1	&	0.552	&	9.66E-12	&	15	&	2440	&	240	&	1	\\	
SDSSJ143123.67+202142.8	&	0.578	&	3.12E-12	&	17	&	2240	&	220	&	1a	 	\\
SDSSJ143633.75+065655.0	&	0.400	&	2.87E-12	&	49	&	1690	&	170	&	1	\\	
SDSSJ144448.25+234554.3	&	0.629	&	3.14E-12	&	17	&	2800	&	280	&	1a	 	\\
SDSSJ144733.05+345506.7	&	0.662	&	8.31E-12	&	41	&	2300	&	230	&	1a	 	\\
SDSSJ145543.45+300322.3	&	0.629	&	2.96E-12	&	18	&	2140	&	210	&	1	\\	
SDSSJ154333.94+102231.9	&	0.487	&	4.40E-12	&	29	&	2250	&	220	&	1	\\	
SDSSJ154823.46+141407.8	&	0.581	&	3.04E-12	&	24	&	1890	&	190	&	1a	 	\\
SDSSJ161924.11+260907.2	&	0.629	&	4.80E-12	&	25	&	2400	&	240	&	1	\\	
SDSSJ162817.12+200348.7	&	0.571	&	3.59E-12	&	16	&	2220	&	220	&	1	\\	
SDSSJ165722.08+395551.4	&	0.579	&	3.01E-12	&	16	&	2400	&	240	&	1	\\	
HE0122-3759 	&	2.200	&	1.44E-11	&	77	&	3400	&	300	&	2	 	\\
HE0359-3959	&	1.521	&	9.18E-12	&	40	&	4250	&	400	&	2	 	\\
HE1003+0149 	&	1.080	&	9.89E-12	&	32	&	2900	&	300	&	2	 	\\
HE1347-2457	&	2.599	&	1.88E-11	&	25	&	6500	 &	600	&	2	 	\\
HE1430-0041	&	1.122	&	8.42E-12	&	40	&	4720	&	300	&	2	 	\\
HE1505+0212	&	1.094	&	1.41E-11	&	81	&	5230	&	200	&	2	 	\\
HE2305-5315 	&	1.073	&	1.68E-11	&	50	&	3300	&	500	&	2	 	\\
\hline\end{tabular}
\begin{list}{}{}
\item[$^\mathrm{a}$]{Redshift provided in the SDSS file header.}
\item[$^\mathrm{b}$]{At 5100 \AA\ rest frame.}
\item[$^\mathrm{c}$]{1 $\sigma$\ confidence level.}
%\item[$^\mathrm{d}$]{S/N at 5100 }
\end{list}
\end{center}
\end{table*}

\clearpage

\hoffset=-1cm
\begin{table*}
\begin{center}
\caption{Measured quantities for the \aliii\ based samples \label{tab:aliii}}
\setlength{\tabcolsep}{5pt}
\begin{tabular}{cccccccc}\hline\hline
\multicolumn{1}{c}{Source} &   $z^{\mathrm{a}}$     &  $\lambda f_{\lambda}^\mathrm{b}$ & S/N & FWHM& $\delta $FWHM$^{\mathrm{c}}$ & Sample   \\ 
& &  [\ergss \cmq] & & [\kms] & [\kms]&\\
\hline 
SDSSJ013514.52-005319.0	&	2.113	&	6.13E-12	&	15	&	3440	&	800	&	3	\\
SDSSJ073149.53+284357.1	&	2.227	&	1.25E-11	&	17	&	4480	&	1100	&	3	\\
SDSSJ075220.38+165506.6	&	2.109	&	7.70E-12	&	24	&	5320	&	900	&	3	\\
SDSSJ081717.14+370252.0	&	2.064	&	1.03E-11	&	23	&	3500	&	400	&	3	\\
SDSSJ082936.31+080140.6	&	2.196	&	1.10E-11	&	21	&	4970	&	500	&	3	\\
SDSSJ084258.83+361444.2	&	2.490	&	8.79E-12	&	22	&	2850	&	400	&	3	\\
SDSSJ084502.73+081214.2	&	2.348	&	7.87E-12	&	15	&	4860	&	600	&	3	\\
SDSSJ084525.84+072222.3	&	2.306	&	1.36E-11	&	21	&	3700	&	900	&	3	\\
SDSSJ085406.12+423810.7	&	2.386	&	6.32E-12	&	16	&	5110	&	600	&	3	\\
SDSSJ091942.82+340301.3	&	2.447	&	1.12E-11	&	20	&	5120	&	900	&	3	\\
SDSSJ093403.96+315331.3	&	2.422	&	2.16E-11	&	30	&	3850	&	900	&	3	\\
SDSSJ094144.72+231144.2	&	2.541	&	1.18E-11	&	22	&	3660	&	900	&	3	\\
SDSSJ094707.81+481613.8	&	2.439	&	1.10E-11	&	20	&	5100	&	900	&	3	\\
SDSS J094748.07+193920.0	&	2.255	&	1.20E-11	&	18	&	5700	&	500	&	3	\\
SDSSJ095707.82+184739.9	&	2.341	&	1.78E-11	&	29	&	3520	&	900	&	3	\\
SDSSJ095817.81+494618.3	&	2.352	&	9.63E-12	&	21	&	4500	&	900	&	3	\\
SDSSJ100356.15-005940.4	&	2.107	&	6.07E-12	&	16	&	4250	&	1100	&	3	\\
SDSSJ100459.14+470058.0	&	2.577	&	1.54E-11	&	29	&	4900	&	900	&	3	\\
SDSSJ100513.60+004028.4	&	2.557	&	1.95E-11	&	18	&	3280	&	300	&	3	\\
SDSSJ104930.88+543839.7	&	2.549	&	1.18E-11	&	19	&	5380	&	500	&	3	\\
SDSSJ105427.17+253600.8	&	2.400	&	1.53E-11	&	23	&	3780	&	900	&	3	\\
SDSSJ111154.35+372321.2	&	2.058	&	1.44E-11	&	25	&	5470	&	900	&	3	\\
SDSSJ112140.13+322346.6	&	2.177	&	1.56E-11	&	22	&	3170	&	300	&	3	\\
SDSSJ114925.65+665949.4	&	2.240	&	9.39E-12	&	29	&	4290	&	900	&	3	\\
SDSSJ120821.01+090130.3	&	2.075	&	1.24E-11	&	24	&	3920	&	900	&	3	\\
SDSSJ122709.48+310749.3	&	2.173	&	1.33E-11	&	19	&	4890	&	900	&	3	\\
SDSSJ125257.96+274542.4	&	2.001	&	6.89E-12	&	19	&	4420	&	500	&	3	\\
SDSSJ125914.85+672011.8	&	2.443	&	1.34E-11	&	25	&	3060	&	400	&	3	\\
SDSSJ125924.28+445105.0	&	2.030	&	1.80E-11	&	28	&	2780	&	400	&	3	\\
\hline\end{tabular}
\begin{list}{}{}
\item[$^\mathrm{a}$]{Redshift provided in the SDSS file header.}
\item[$^\mathrm{b}$]{At 1800 \AA\ rest frame.}
\item[$^\mathrm{c}$]{1 $\sigma$\ confidence level.}
\end{list}
\end{center}
\end{table*}

\clearpage

\addtocounter{table}{-1}

\hoffset=-1cm
\begin{table*}
\begin{center}
\caption{(cont.) Measured quantities for the \aliii\ based samples}
\setlength{\tabcolsep}{5pt}
\begin{tabular}{lccccccc}\hline\hline
\multicolumn{1}{c}{Source} &   $z^{\mathrm{a}}$     &  $\lambda f_{\lambda}^\mathrm{b}$ & S/N & FWHM& $\delta $FWHM$^{\mathrm{c}}$ & Sample    \\ 
& &  [\ergss \cmq] & & [\kms] & [\kms]&\\
\hline 
SDSSJ130236.17+095831.8	&	2.073	&	1.12E-11	&	22	&	5170	&	900	&	3	\\
SDSSJ130924.78+412427.9	&	2.081	&	9.08E-12	&	17	&	2500	&	800	&	3	\\
SDSSJ132615.16-030357.5	&	2.121	&	2.45E-11	&	33	&	5220	&	900	&	3	\\
SDSSJ142402.65+263624.8	&	2.095	&	6.44E-12	&	17	&	5270	&	600	&	3	\\
SDSSJ142500.24+494729.3	&	2.253	&	2.24E-11	&	24	&	4050	&	900	&	3	\\
SDSSJ153837.01+522555.1	&	2.078	&	8.34E-12	&	18	&	3630	&	600	&	3	\\
SDSSJ154624.57-014849.2	&	2.009	&	8.15E-12	&	18	&	3050	&	800	&	3	\\
SDSSJ154757.71+060626.6	&	2.018	&	1.82E-11	&	35	&	4300	&	900	&	3	\\
SDSSJ160955.41+065401.9	&	2.134	&	1.06E-11	&	22	&	4910	&	900	&	3        \\
SDSSJ162115.06+400732.2	&	2.206	&	1.21E-11	&	47	&	2390	&	300	&	3	\\
SDSSJ210831.56-063022.5	&	2.345	&	2.53E-11	&	61	&	5060	&	500	&	3	\\
SDSSJ223304.01-080142.6	&	2.264	&	8.82E-12	&	17	&	4160	&	1100	&	3	\\
SDSSJ232115.48+142131.5	&	2.537	&	1.05E-11	&	16	&	4260	&	1100	&	3	\\
\hline\end{tabular}
\begin{list}{}{}
\item[$^\mathrm{a}$]{Redshift provided in the SDSS file header.}
\item[$^\mathrm{b}$]{At 1800 \AA\ rest frame.}
\item[$^\mathrm{c}$]{1 $\sigma$\ confidence level.}
\end{list}
\end{center}
\end{table*}

\begin{table*}
%\tabletypesize{\scriptsize}
%\rotate
%\tablewidth{230pt}
\caption{Results from Preliminary Sample \label{tbl-1}}
%&&\multicolumn{2}{c}{$H_{0}$=60} &    & \multicolumn{2}{c}{$H_{0}$=70} &&\multicolumn{2}{c}{$H_{0}$=80} \\ \cline{3-4} \cline{6-7} \cline{9-10}
%&& }%
\begin{tabular}{lccccc}
\hline\hline
Selected Cosmologies	\\
&& \multicolumn{1}{c}{$\bar{\Delta}/\sigma_{\bar{\Delta}}^a$}  &  \multicolumn{1}{c}{$b^b$  }& \multicolumn{1}{c}{$\chi^2~^c$}\\ \hline
%\cline{1-6}
%\\ 
Concordance	&	 &		1.21	 &   0.03 &	1.05	\\        
$\Lambda$-dominated & &	 		-6.53& -0.18 &	1.67	\\
M-dominated	&		& -0.12	& 0.125 &	1.84 	\\
Little Matter	&	          &  1.74 	& -0.03 &	1.07	\\
Empty     &	 		&0.64	& 0.06& 	1.06	\\
\hline
\end{tabular}
\begin{list}{}{}
\item[$^\mathrm{a}$]{Ratio between the average of $\delta \log L$ and the average standard deviation.}
\item[$^\mathrm{b}$]{Slope of linear fit. For our sample, $\sigma_{b}\approx 0.075$. }
\item[$^\mathrm{c}$]{Normalized $\chi^{2}$.}
\end{list}
\end{table*}

\clearpage

\begin{table*}
%\tabletypesize{\scriptsize}
%\rotate
\caption{Error Budget\label{tab:budget}}
\begin{tabular}{lcccc}
%\tablewidth{0pt}
\hline\hline
Parameter $p$ &  $\delta\log p^a$  & Power & \\  \hline
\\
&\multicolumn{4}{c}{Virial luminosity}\\
\\
$\lambda_\mathrm{Edd}$	&	0.13	 	&	2	\\
$\kappa/(\bar{\nu_\mathrm{i}})$	&	0.020 -- 0.033	 	&	1	\\
$10^{n_\mathrm{H}U}$ &  0.050 -- 0.100   & 1 \\
$f_\mathrm{S}$	&	0.043 -- 0.087	 		&	2	\\
FWHM	&		 	0.065	&	4	\\
\\
Prop. err.	&	0.379 -- 0.418	  	&		\\
	&		 		&		\\
&\multicolumn{4}{c}{$z$-based luminosity}\\
&& \\
$f_\lambda$	&	0.043		&	1	\\
$z$	&		0.000 -- 0.001	&	2	\\
$B. C.$	&	    0.043 -- 0.087	 		&	1	\\
Anisotropy &        0.085 -- 0.15 & 1 \\
\\
Prop. err.	&  0.105 --	 0.179  	&		\\
	&		&		&		\\
Total err.	&	 	 	0.394 -- 0.455	&		\\
%\tablecomments{Table \ref{tbl-1} is published in its entirety in the  electronic edition of the {\it Ast.rophysical Journal}.  A portion is  shown here for guidance regarding its form and content.}
%\tablenotetext{a}{Slope of best fitting line (unweighted $\chi^{2}$) of $\delta \log L$ vs. $z$. Uncertainty is $\pm$ 0.07 (standard error propagation) and $\pm$ 0.07 (bootstrap). The slope and its uncertainty are not dependent on $H_{0}$. }
%\tablenotetext{b}{Ratio between the average of $\delta \log L$ and the average standard deviation.}
 \hline
 \end{tabular}
 \begin{list}{}{}
\item[$^\mathrm{a}$]{Estimated statical errors for the actual sample of 92 sources presented in this paper. }
%\item[$^\mathrm{b}$]{Statistical errors predicted for a sample with reduced statistical uncertainties; see text for details.}
\end{list}
\end{table*}

%\end{document}
\clearpage

\clearpage

%\begin{figure}
%\epsscale{1.00}
%\includegraphics{fig04.eps}
%\caption{Left panel: behaviour of \lledd\  for the sources in the sample extracted from \citet{shenetal11} as a function of \rfe\ (in linear scale). The dashed and the dotted lines show the limiting criteria on \feii\ for the definition of four subsamples  in the right panels. %The thick red lines shows the convolution of a function modelling orientation effects and a log-normal distribution of \lledd\ with $\sigma = 0.2$. See text for details.  
%\label{fig:lledd}}
%\end{figure}

\clearpage

\clearpage

\clearpage

%\begin{figure}
%\epsscale{.}
%\plotone{slope.eps}
%\caption{Slope of residuals between virial estimates and models, for a flat universe
% as a function of $H_{0}$ and $\Omega_\mathrm{M}$. \label{fig4}}
%\end{figure}
\clearpage

\eject
\clearpage

%% Here we use \plottwo to present two versions of the same figure,
%% one in black and white for print the other in RGB color
%% for online presentation. Note that the caption indicates
%% that a color version of the figure will be available online.
%%

%% If you are not including electonic art with your submission, you may
%% mark up your captions using the \figcaption command. See the
%% No more than seven \figcaption commands are allowed per page,
%% so if you have more than seven captions, insert a \clearpage

%% Tables should be submitted one per page, so put a \clearpage before

%% Two options are available to the author for producing tables:  the
%% deluxetable environment provided by the AASTeX package or the LaTeX
%% table environment.  Use of deluxetable is preferred.
%%

%% Three table samples follow, two marked up in the deluxetable environment,
%% one marked up as a LaTeX table.

%% In this first example, note that the \tabletypesize{}
%% command has been used to reduce the font size of the table.
%% We also use the \rotate command to rotate the table to
%% landscape orientation since it is very wide even at the
%% Note also that the \label command needs to be placed

%% This table also includes a table comment indicating that the full
%% version will be available in machine-readable format in the electronic

\clearpage

\end{document}